\documentclass[prb,twocolumn,nofootinbib,superscriptaddress,aps,amsmath,amsfonts,notitlepage,longbibliography]{revtex4-1}
\usepackage{amsmath,amsfonts,amssymb,dsfont,graphicx,bm,caption,subcaption}
\graphicspath{{../../Figures/}}
\usepackage{color}
\usepackage{hyperref}
\usepackage{tikz}
\usepackage{pbox}
\usepackage{ulem}

\usetikzlibrary{calc}

\def\be{\begin{equation}}
\def\ee{\end{equation}}
\def\bea{\begin{eqnarray}}
\def\eea{\end{eqnarray}}
\def\bpm{\begin{pmatrix}}
\def\epm{\end{pmatrix}}

\def\Im{\mathop{\rm Im}}
\def\Re{\mathop{\rm Re}}

\def\Tr{\mathop{\rm Tr}}
\def\sign{\mathop{\rm sign}}

\newcommand{\e}{\epsilon}

\newcommand{\p}{\partial}

\newcommand{\ket}[1]{\left| #1\right\rangle}

\begin{document}
\title{Theory of weak symmetry breaking of translations in $\mathbb{Z}_2$ topologically ordered states and its relation to topological superconductivity from an exact lattice $\mathbb{Z}_2$ charge-flux attachment}
\author{Peng Rao}
\affiliation{Max Planck Institute for the Physics of Complex Systems, N\"othnitzerstr. 38, 01187 Dresden, Germany}
\author{Inti Sodemann}
\affiliation{Max Planck Institute for the Physics of Complex Systems, N\"othnitzerstr. 38, 01187 Dresden, Germany}
\affiliation{Department of Physics and Astronomy, University of California, Irvine, California 92697, USA}
\date{\today}
\begin{abstract}
We study $\mathbb{Z}_2$ topologically ordered states enriched by translational symmetry by employing a recently developed 2D bosonization approach that implements an exact $\mathbb{Z}_2$ charge-flux attachment in the lattice. Such states can display `weak symmetry breaking' of translations, in which both the Hamiltonian and ground state remain fully translational invariant but the symmetry is `broken' by its anyon quasi-particles, in the sense that its action maps them into a different super-selection sector. We demonstrate that this phenomenon occurs when the fermionic spinons form a weak topological superconductor in the form of a 2D stack of 1D Kitaev wires, leading to the amusing property that there is no local operator that can transport the $\pi$-flux quasi-particle across a single Kitaev wire of fermonic spinons without paying an energy gap in spite of the vacuum remaining fully translational invariant. We explain why this phenomenon occurs hand-in-hand with other previously identified peculiar features such as ground state degeneracy dependence on the size of the torus and the appearance of dangling boundary Majorana modes in certain $\mathbb{Z}_2$ topologically ordered states. Moreover, by extending the $\mathbb{Z}_2$ charge-flux attachment to open lattices and cylinders, we construct a plethora of exactly solvable models providing an exact description of their dispersive Majorana gapless boundary modes. We also review the $\mathbb{Z}\times (\mathbb{Z}_2)^3$ classification of 2D BdG Hamiltonians (Class D) enriched by translational symmetry and provide arguments on its robust stability against interactions and self-averaging disorder that preserves translational symmetry.
\end{abstract}

\maketitle

\section{Introduction}\label{Sec:Introduction}

The Toric Code (TC)~\cite{Kitaev2003} is a simple example of an exactly solvable model of $\mathbb{Z}_2$ topologically ordered states~\cite{Wen2007,Fradkin2013}. But more than providing a single clear example of these remarkable states, it offers a new set of building blocks to construct a plethora of other states.~\cite{Oscar2020} These building blocks are its non-trivial quasiparticles $e, m$ and $\varepsilon$. $e$ and $m$ are hard-core bosons and $\varepsilon$ is a fermion, and they all see each other as semions (`$\pi$-fluxes'). One can describe any state of the physical Hilbert space in a basis in which one keeps track of the occupations of only two of these particles, since one of them can always be viewed as the bound state of the other two.~\cite{Chen2018,Oscar2020}

\begin{figure}
	
	\includegraphics*[width=0.8\linewidth]{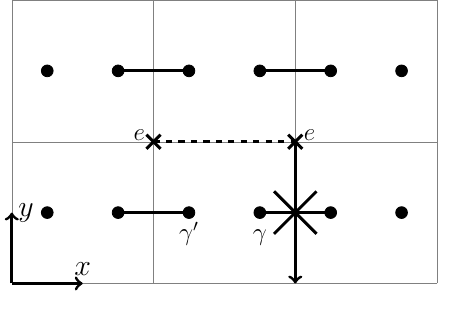}
	\caption{\label{Fig:AnyonHopping} Depiction of $\mathbb{Z}_2$ topologically ordered state with weak symmetry breaking along $y$-direction, where the $\varepsilon$-particles form a stack of Kitaev wires along the $x$-direction. Solid lines depict the ground state pairing of $\varepsilon$ Majorana modes (black dots). There are dangling modes at the boundary. The $e$-particles on vertices (small cross) can hop along the $x$-direction (dashed line), but there is no local operator that hops them in the $y$-direction across a single wire without paying the Bogoliubov fermion gap in spite of this being a symmetry.}
\end{figure}

Importantly, these particles are non-local: they can only be created in pairs at the open ends of certain operator strings. Therefore, any physical state must respect the parity conservation of these particles. These parity symmetries are a kind of `tautology', in an analogous sense to how an open string always necessarily has two ends. Therefore, these symmetries can never be broken \textit{explicitly} by any terms added to the Hamiltonian. Remarkably, however, since these parity symmetries are global, they can be broken \textit{spontaneously}. This occurs, for example, by adding a finite density of one of the bosonic particles (say $m$) to the TC vacuum and having it form a Bose-Einstein condensate.~\cite{Wen2007,Fradkin2013} Such phases in which the unbreakable parity symmetry is spontaneously broken, correspond to trivial short ranged-entangled phases. This is intimately related to the long-range phase rigidity of this condensate, leading to energetically costly long-ranged distortions for inserting the anyon that is seen as a $\pi$-flux by the condensate. On the other hand, when a finite density of the bosonic anyons are added to the TC vacuum but instead they form an `atomic insulator' state in which they are localized at sites without spontaneously breaking their parity symmetry, the resulting state is still $\mathbb{Z}_2$ topologically ordered, although it can display a projective symmetry implementation of the translation group.~\cite{Wen2002,Wen20021}

However, adding the $\varepsilon$-fermions onto the TC vacuum affords much more flexibility in constructing non-trivial states. If $e$-particles are kept dynamically immobile, these constructions can be viewed as a form of $\mathbb{Z}_2$ charge-flux attachment implementing a type of local 2D Jordan-Wigner transformation.~\cite{Bravyi2002,Levin2003,Verstraete2005,Ball2005,Levin2006,Gaiotto2016,Chen2018,Radicevic2019,Chen2019,Chen20191,Oscar2020,Borla2020} In this case, and in contrast to the bosonic case, any local fermion Hamiltonian always respects parity. Therefore the state lacks any form of long-range parity-phase rigidity, and distant immobile anyons ($e$-particles) that are seen as a $\pi$-fluxes by the fermions can be inserted with a finite energy cost. In fact the celebrated Kitaev honeycomb model~\cite{Kitaev2006} can be viewed as a special case of this construction,~\cite{Chen2018} and deconfinement of the $\pi$-fluxes in these states with a finite density of $\varepsilon$-fermions remains even when they form a gapless Fermi sea~\cite{Oscar2020} akin to an orthogonal metal.~\cite{Nandkishore2012} For other studies of local boson-fermion mappings, see also Refs.~\onlinecite{Kitaev2006,Chen2007,Chen2008,Cobanera2010,Cobanera2011,Nussinov2012}.

Even though the fermion parity symmetry cannot be broken spontaneously in the proper sense, the 1D topological phase of a Kitaev wire has certain features resembling spontaneous parity symmetry breaking.~\cite{Kitaev2001} In this study, we will demonstrate how states containing such Kitaev wires of the emergent $\varepsilon$-fermions underlie a remarkable phenomenon dubbed `weak symmetry
breaking' in the case of translational symmetry in $\mathbb{Z}_2$ topologically
ordered states.~\cite{Kitaev2006} A state weakly breaking translational symmetry is one in which its ground state is exactly translationally invariant, but the symmetry is in a sense broken by its anyon quasi-particles. To be precise, it is the situation in which the symmetry action on its anyon quasi-particles cannot be implemented locally and maps them between different super-selection sectors;~\cite{Barkeshli2019} this phenomenon was also referred to as `unconventional' symmetry implementation in Ref.~\onlinecite{Lu2016}. The reason for the appearance of weak symmetry breaking in stacks of Kitaev wires of $\varepsilon$-fermions, is related to the fact that such wires display a `locking' of fermion parity and boundary conditions twist, namely, their ground state has an odd (even) number of fermions for periodic (anti-periodic) boundary conditions. As a consequence, if a $\pi$-flux crosses a Kitaev wire, it will swap the boundary condition of the wire, and such operations would necessarily excite a single Bogoliubov fermion above the gap, as depicted in Fig.~\ref{Fig:AnyonHopping}. However, it is impossible to remove such a single fermion by any local operation, because local operations can only add or remove fermions in pairs. Therefore, the $\pi$-flux cannot be transported to any site in which it crosses an odd number of $\varepsilon$-fermion wires even though such sites are related by translational symmetry (see Fig.~\ref{Fig:AnyonHopping}). As a consequence these states will display two types of fluxes belonging to two superselection sectors.

Our work builds on a series of several key previous studies. These anomalies of the implementation of translational symmetry have been investigated by a series of works in the past,~\cite{Wen2003,Kitaev2006,Kou2008,Wen2009,Bombin2010,Wen2010,Cho2012,Jing2013} where it was emphasized that $\mathbb{Z}_2$ topologically ordered states can have a size dependent ground state degeneracy (GSD) in the torus different from $4$, and display features such as edge dangling Majorana modes protected by translational symmetry. The Wen plaquette model was the first and seminal example of such states.~\cite{Wen2003} We will combine this understanding with the recently completed classification of 2D topologically superconductors enriched by translational symmetry~\cite{Wen2009,Ryu2010,Sato12010,Wen2010,Sato2010,Essin2013,Mesaros2013,Qi2013,Yao2013,Wang2014,Metlitski2014,Ono2020,Geier2020,Schindler2020,Ono2021}, exploiting the exact lattice $\mathbb{Z}_2$ charge-flux attachment,~\cite{Chen2018} to develop an overarching picture of the interplay of translational symmetry and $\mathbb{Z}_2$ topological order. In particular, we will be able to specify when a state will have a projective symmetry implementation and when the symmetry will be weakly broken for any topological paired state of $\varepsilon$-fermions with translational symmetry. We will then link the appearance of dangling boundary Majorana modes with the existence of stacks of Kitaev wires and the bulk weak symmetry breaking of translations of fluxes. In doing so, we will extend the constructions of Refs.~\onlinecite{Chen2018,Oscar2020} to lattices with fully open boundaries and cylinders and provide exactly solvable models for the bulk and edge excitations. We note that, because translational symmetry swaps the super-selection sectors of the anyons in states with weak symmetry breaking, this phenomenon is beyond the projective symmetry group construction,~\cite{Wen2002,Wen20021} and also beyond the considerations of Ref.~\onlinecite{Essin2013}. Also, since translational symmetry is not exactly on-site, it is also beyond the considerations of Ref.~\onlinecite{Mesaros2013}. We also note in passing that a related form of weak symmetry breaking of translations in fractional quantum Hall states has been recently studied in Ref.~\onlinecite{Tam2020}.

Since our paper is quite lengthly we have provided a succinct summary of main results in the Sec.~\ref{Sec:Conclusions}, which can be read in an essentially independent way of the main body of the paper. The remainder of the paper is organized as follows. In Section~\ref{Sec:PeriodicLattice} we extend this construction to lattices with open boundaries. In Section~\ref{Sec:TopologicalSuperconductor} we review the classification and bulk-boundary correspondence of 2D BdG Hamiltonians with translational symmetry. In Section~\ref{Sec:WeakSymmetry} we apply this machinery to develop a theory of the lattice-size-dependent ground state degeneracy, the dangling Majorana modes, and the weak symmetry breaking of translations of $\mathbb{Z}_2$ topologically ordered states. In section~\ref{Sec:Model} we write down and analyze an exactly solvable model that interpolates from the TC to the Kitaev honeycomb model and realizes many examples of the aforementioned properties of translationally symmetric $\mathbb{Z}_2$ topologically ordered states. Several technical aspects and alternative derivations are presented in Appendices~\ref{Appendix:Duality}-\ref{Appendix:ChernNumber}.
\begin{figure*}
	
	\includegraphics*[width=0.3\linewidth]{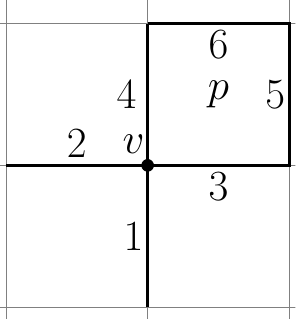}\llap{
		\parbox[b]{10cm}{\large (a)\\\rule{0ex}{4.9cm}
	}}~~
	\includegraphics*[width=0.33\linewidth]{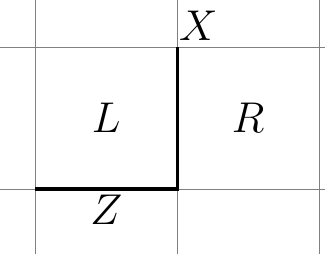}\llap{
		\parbox[b]{11.3cm}{\large (b)\\\rule{0ex}{4.9cm}
	}}~~
	\includegraphics*[width=0.25\linewidth]{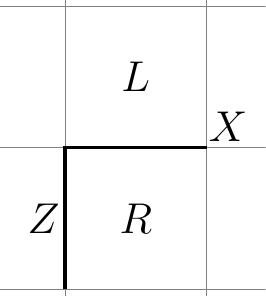}	\llap{
		\parbox[b]{8.8cm}{\large (c)\\\rule{0ex}{4.9cm}
	}}
	\caption{\label{Fig:OperatorDefinitions} (a) Representation of $\Gamma^e_v, \Gamma^\varepsilon_p$ and $U_{p,x,y}$ defined on vertex $v$ and plaquette $p$. Here spins reside in the links of the square lattice, and those participating in these operators are shown as solid black lines. (b)-(c) Definitions of $L$ and $R$ plaquettes for the mapping Eq.~(\ref{Eqn:OperatorDefinitions}).}
\end{figure*}

\section{Representation of Particles in Toric Code} \label{Sec:Review}

In this work we would like to advance the point of view that the TC Hamiltonian provides an exact re-writing of a Hilbert space of local degrees of freedom in terms of non-local degrees of freedom. These local or physical degrees of freedom are spin-$1/2$, or equivalently hard-core bosons, residing in the links of a square lattice. In its traditional formulation, the non-local or unphysical degrees of freedom can be viewed also as spin-$1/2$ residing in the vertices and the plaquettes. More specifically, the states of such non-local degrees of freedom are labeled by the $\pm 1$ eigenvalues of operators $G_v^e$ and $G_p^m$, defined on each vertex $v$ and plaquette $p$:
\be\label{Eqn:TCOperators}
G^e_v = X_3X_4X_1 X_2,~G^m_p= Z_3Z_5Z_6 Z_4;
\ee
where the convention is depicted in Fig.~\ref{Fig:OperatorDefinitions}. When placed in a torus such operators satisfy a global constraint:
\be\label{Eqn:Constraints1}
\prod_{p } G^m_p=1,~ \prod_{v } G^e_v=1,
\ee
where the product is taken over all plaquettes and vertices in the lattice. More specifically, we say that when $G^e_v=-1$ ($G^m_p=-1$) an $e$ ($m$) hard-core bosonic particle resides in the corresponding vertex (plaquette). In order to account for the above constraint of Eq.~(\ref{Eqn:Constraints1}) in the torus, we take these non-local hard-core bosonic particles to satisfy separate global $\mathbb{Z}_2$ number parity conservation symmetries, and we would only interpret parity even subspaces as physical, and discard all the states with a total odd number of hard-core bosons as unphysical. The non-locality of these bosonic degrees of freedom stems from the fact that any Hamiltonian which is local in the underlying local physical spins degrees of freedom maps onto a Hamiltonian in which the $e$ and $m$ bosons experience a non-local mutual semionic statistical interaction.~\cite{Kitaev2003,Kitaev2006} Hamiltonians in which one of the boson species is held immobile while the other is allowed to hop and pair fluctuate on top of the TC vacuum are examples of classic bosonic $\mathbb{Z}_2$ lattice gauge theories.~\cite{Wen2007} Each subspace of such Hamiltonians is labeled by the static location of the immobile particles, while the remaining mobile particles can be viewed as ordinary hard-core bosons moving in a background configuration of static $\pi$-fluxes.~\cite{Oscar2020}

More recently a different re-writing of the microscopic Hilbert space in terms of other non-local degrees of freedom has been introduced in Ref.~\onlinecite{Chen2018}. For related ideas and elaborations see also Refs.~\onlinecite{Bravyi2002,Levin2003,Verstraete2005,Ball2005,Levin2006,Gaiotto2016,Radicevic2019,Chen20191,Oscar2020}. The idea behind this construction is to exploit the property that the bound state of the $e$ and $m$ particles, denoted by $\varepsilon$, has fermionic exchange statistics relative to itself, and therefore can be used to introduce a non-local degree of freedom that is a fermion, rather than hard-core boson. Therefore, rather than using $e$ and $m$ as a basis, we can alternatively represent exactly the entire Hilbert space associated with any local spin Hamiltonian by introducing an $\varepsilon$ spinless complex fermion (two Majorana modes) residing in the plaquettes, and an $e$ hard-core boson residing at the vertices.~\cite{Oscar2020} (see Fig.~\ref{Fig:AnyonHopping}) In this new representation, the operator that used to measure the parity of the $m$ boson is now taken to measure parity of the $\varepsilon$-fermion:
\be\label{Eqn:FermionParity}
\Gamma^\varepsilon_p = Z_3Z_5Z_6 Z_4.
\ee
Therefore, we say that an $\varepsilon$ fermion resides in the plaquette $p$ if $\Gamma^\varepsilon_p=-1$. On the other hand the operator measuring the parity of the $e$ boson is now replaced by a new composite operator, which requires a pairing convention for plaquettes and vertices, which we do so following the convention of Ref.~\onlinecite{Chen2018}, by pairing each vertex with its North-East plaquette, as depicted in Fig.~\ref{Fig:OperatorDefinitions}, and the $e$-parity is defined as:
\be\label{Eqn:eDefinition}
\Gamma^e_v =  X_3X_4X_1 X_2 \times Z_3Z_5Z_6 Z_4.
\ee

\begin{figure}
		\includegraphics*[width=0.55\linewidth]{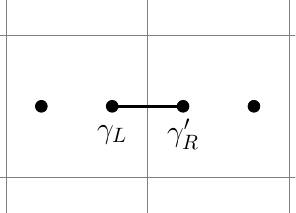}\llap{
			\parbox[b]{8.8 cm}{\large (a)\\\rule{0ex}{3.7cm}
		}}~
		\includegraphics*[width=0.35\linewidth]{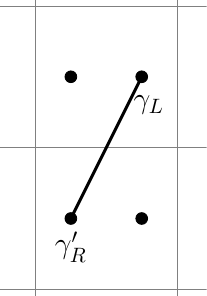}\llap{
			\parbox[b]{5.6cm}{\large (b)\\\rule{0ex}{3.7cm}
		}}
	\caption{\label{Fig:MajoranaHop} Majorana representation of (a) horizontal, (b) vertical $\varepsilon$-hopping in Eq.~(\ref{Eqn:OperatorDefinitions}).}
	
\end{figure}

\begin{figure*}
	
	\includegraphics*[width=0.4\linewidth]{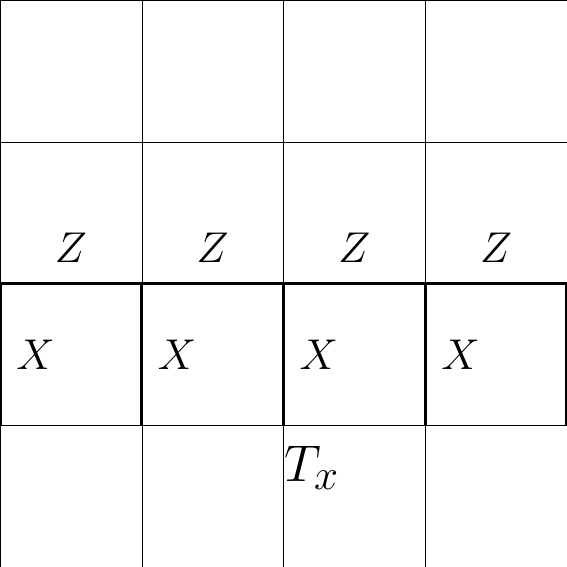}\llap{
		\parbox[b]{13.5cm}{\large (a)\\\rule{0ex}{6.5cm}
	}}~~
	\includegraphics*[width=0.4\linewidth]{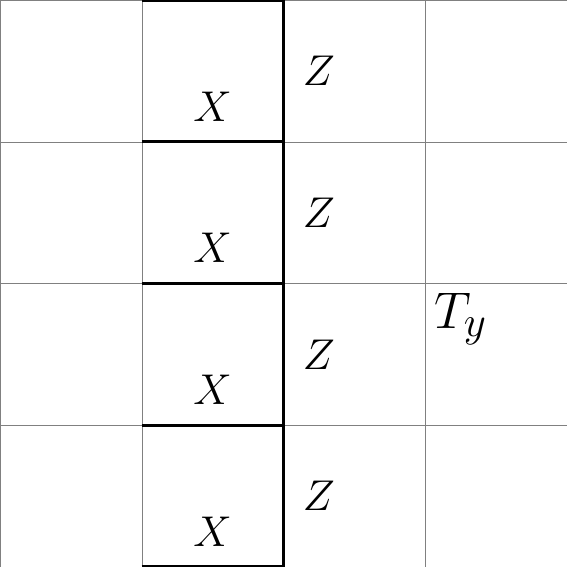}\llap{
		\parbox[b]{13.5cm}{\large (b)\\\rule{0ex}{6.5cm}
	}}~~
	
	\caption{\label{Fig:TopologicalDegeneracies} (a) Representation of the t'Hooft operator $T_x$ along the $x$-direction. (b) Visual representation of the t'Hooft operator $T_y$ along the $y$-direction. The lattice size is $L_x=L_y=4$. }
\end{figure*}
Similarly, we say that an $e$ hard-core boson resides in a vertex $v$ if $\Gamma^e_v=-1$. The current rewriting allows to represent the local Hamiltonians of the microscopic spins in terms of Hamiltonians for the $\varepsilon$-fermion and the $e$ boson which experience a non-local mutual semionic interaction. If the $e$-particles are held immobile by enforcing that all operators in the Hamiltonian commute with the local $e$-particle number, $\Gamma^e_v$ for all vertices of the lattice, the resulting theory can be viewed as a modified $\mathbb{Z}_2$ lattice gauge theory, whose gauge invariant subspaces correspond to those of ordinary fermion Hamiltonians subjected to non-dynamical static background $\pi$ magnetic flux tubes at the vertices that contain an $e$ boson.~\cite{Oscar2020} In particular, the subspace without flux ($\Gamma^e_v=1$ for all vertices) can be viewed as an ordinary fermionic Hilbert space, and thus the restriction to this subspace is a systematic form of local higher dimensional bosonization of fermion models.~\cite{Chen2018}

Before describing finite size geometries we will review this fermionic representation in the infinite plane following the convention from Ref.~\onlinecite{Chen2018}. We define two elementary $\varepsilon$ pair-creation operators as follows:
\be\label{Eqn:PairCreation}
U_{x,p} = X_5 Z_3,~U_{p,y} = X_6Z_4,
\ee
that create a pair of $\varepsilon$ particles on plaquette $p$ and its nearest neighbour to its East and North, as shown in Fig.~\ref{Fig:OperatorDefinitions}. Together with $\Gamma^\varepsilon_v$, they form a complete algebraic basis of spatially local operators out of which any operator that commutes with all $\Gamma^\e_v$ from Eq.~(\ref{Eqn:eDefinition}) can be obtained by multiplying and adding these. These operators can therefore be mapped exactly to a complete set of parity-even fermionic operators in a way that preserves space locality. To do so we introduce two Majorana fermion operators in every plaquette, $\gamma_p$ and $\gamma_p'$, and map their bilinear products onto operators acting on the underlying physical spins as follows: (see Fig.~\ref{Fig:MajoranaHop})
\be\label{Eqn:OperatorDefinitions}
U_{p}  \rightarrow i\gamma_L \gamma'_R,~\Gamma^\varepsilon_p \rightarrow -i \gamma_p\gamma'_p.
\ee
Directionality $L, R$ follows the same convention as in Ref.~\onlinecite{Chen2018}. The above representation is exact in the subspace where there are no $e$ particles, namely for $\Gamma^e_v=1$ on every vertex $v$, but can be easily extended to cases where there are static $e$-particles.~\cite{Oscar2020} $\gamma, \gamma'$ are related to the  $\varepsilon$-particle complex fermion operator $a$ by:
\be\label{Eqn:MajoranaFermions}
\gamma = a+a^\dagger,~\gamma'=-i(a-a^\dagger).
\ee
We reiterate that this mapping (\ref{Eqn:OperatorDefinitions}) preserves spatial locality in the dual fermionic theory, namely that local spin operators that commute with Eq.~(\ref{Eqn:eDefinition}) are mapped into local fermion operators and it is, therefore, a two-dimensional version of the Jordan-Wigner transformation which preserves locality.

\subsection{Torus Geometry}\label{Sec:TorusGeometry}

We will now generalize the construction of Ref.~\onlinecite{Chen2018} to a finite-size torus with side length $L_x$ and $L_y$ (the lattice constants are taken unity). We begin by describing how to recover the full dimensionality of the underlying Hilbert space of physical spins, which is $2^{2L_x L_y}$, in terms of the dual fermionic $\varepsilon$ and the static bosonic $e$ degrees of freedoms. Since, the $e$ particles are held immobile by enforcing that every operator in the Hamiltonian commutes with $\Gamma^e_v$ from Eq.~(\ref{Eqn:eDefinition}), the Hilbert space decomposes into a direct sum of decoupled subspaces with specific values $\Gamma^e_v=\pm1$. In the torus there are $2^{L_x L_y-1}$ such independent values, since the $\Gamma^e_v$ operators also satisfy a parity constraint:
\be\label{Eqn:eConstraint}
\prod_v \Gamma^e_v=1.
\ee
Notice that if we take the product of $\Gamma^e_v$ over all the vertices contained inside a simply connected region in the torus, one obtains a closed loop operator that acts only on spins at the boundary of such region, which can be viewed as a $\mathbb{Z}_2$ lattice version of the Gauss-Ostrogradsky’s divergence theorem. Clearly such boundary operator must commute with any Hamiltonian, since the Hamiltonian commutes with every $\Gamma^e_v$. However, notice that when such region is not simply connected but wraps around either the $x$ or $y$ directions of the torus, there are two disconnected loop operators that make up the boundary of the region and which wind completely around either of the directions of the torus, as depicted in Fig.~\ref{Fig:TopologicalDegeneracies}. We call these two operators along the $x, y$ directions $T_{x,y}$, and write them explicitly as:
\be\label{Eqn:t'Hooft1}
T_{x,y} = -\prod XZ,
\ee
where the convention for taking the product is depicted in Fig.~\ref{Fig:TopologicalDegeneracies}, and we have added a global minus sign for future notational convenience. Notice that the $T_{x,y}$ operators cannot be expressed in terms of the $\Gamma^e_v$ and therefore they are algebraically independent. Importantly, any local Hamiltonian that commutes with every $\Gamma^e_v$ must also commute with $T_{x,y}$. The spectrum of these operators is $T_{x,y}=\pm1$ , they also commute $[T_x,T_y]=0$, and therefore we have $2^{L_x L_y+1}$ decoupled sectors of the Hilbert space labeled by $\{\Gamma^e_v, T_x, T_y\}$.

Each of these $2^{L_x L_y+1}$ subspace labeled by $\{\Gamma^e_v, T_x, T_y\}$ can be mapped exactly into the parity-even subspace of a Fermionic model with static background $\pi$-fluxes. This parity even restriction appears in the torus because of the constraint of the operator $\Gamma^\varepsilon_p$:
\be\label{Eqn:Constraints2}
\prod \Gamma^\varepsilon_p=1.
\ee
Therefore, in analogy to the bosonic case, we only interpret the parity even subspaces of the fermions as physical and discard all of the states with a total odd number of fermions as unphysical. Since there are $L_x L_y$ plaquettes, this leads to a degeneracy $2^{L_xL_y-1}$for each of these parity-even fermion sub-spaces. As we see, then the total dimensionality of the Hilbert space is recovered from the $2^{L_x L_y+1}$ subspaces labeled by $\{\Gamma^e_v, T_x, T_y\}$, each containing only even numbers of $\varepsilon$-fermions.

\begin{figure}
	
	\includegraphics*[width=\linewidth]{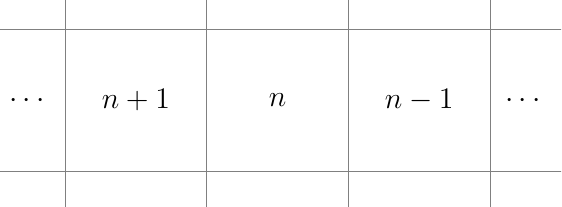}
	
	\caption{\label{Fig:BoundaryTwist} Transport of fermion across a given row given by Eq.~(\ref{Eqn:Consistency}), with the order of product $n$ shown explicitly.}
	
\end{figure}

\begin{figure}
	
	\includegraphics*[width=0.8\linewidth]{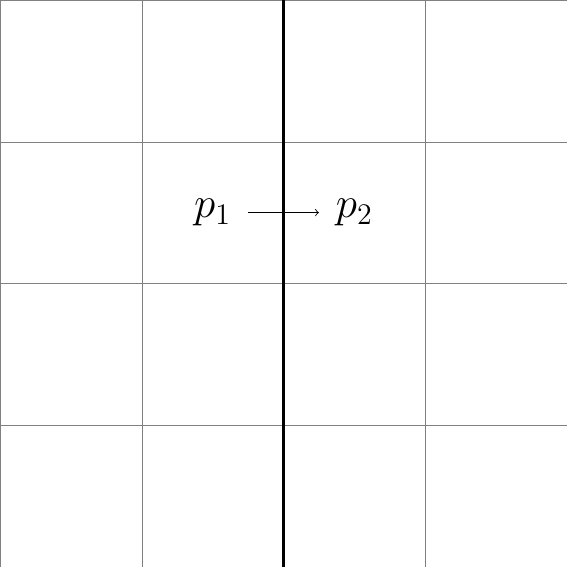}
	
	\caption{\label{Fig:BranchCut} Twist of the horizontal boundary as a branch-cut shown by the bold black line. Fermion transport across the branch-cut has an additional factor of $-1$ in Eq.~(\ref{Eqn:OperatorDefinitions}). For example, $U_{x,p_1}$ is mapped into $-i\gamma_{p_1}\gamma_{p_2}'$.
		}
	
\end{figure}

Now, however, the representation from Eq.~(\ref{Eqn:OperatorDefinitions}) only applies to the sector in which $\Gamma^e_v=1$, and $T_x=T_y=1$, and needs to be modified in other sectors. To show this, we will describe the correspondence between the representation of these operators and the four sectors with arbitrary values of $\{T_x, T_y\}$, but restricted to $\Gamma^e_v=1$; the representation of sectors with $\Gamma^e_v\ne1$ is discussed in Ref.~\onlinecite{Oscar2020}. To do this, notice that the $T_{x,y}$ operators can be written as a string of products of the $U_{x,p}, U_{y,p}$ and $\Gamma^\varepsilon_p$ operators as follows:
\be\label{Eqn:Consistency}
\prod_{n\in \gamma_x }\bigg(\Gamma^\varepsilon_n U_{x,n} \bigg)=T_x,~\prod_{n \in \gamma_y} \bigg(\Gamma^\varepsilon_n U_{y,n} \bigg)=T_y,
\ee
where the product is taken along horizontal and vertical paths $\gamma_{x,y}$ from East to West and South to North respectively. As an example, the convention for $\gamma_x$ in the strings is shown in Fig.~\ref{Fig:BoundaryTwist}. These string operators in Eq.~(\ref{Eqn:Consistency}) can be viewed as the operators associated with the transport of fermions around the non-contractible loops of the torus oriented along $x$- and $y$-directions. Substituting Eq.~(\ref{Eqn:OperatorDefinitions}) in the right-hand side of both equalities in Eq.~(\ref{Eqn:Consistency}) gives $T_{x,y}=1$. Therefore, the subspace with $T_x=T_y=1$ corresponds to fermions having periodic boundary conditions along both directions. The subspaces with $T_x=-1$ $(T_y=-1)$ can be represented as fermions having anti-periodic boundary conditions along the $x$- ($y$-)direction. For example, if $T_x=-1$ and $T_y=1$, we can represent the $U_{x,p}, U_{y,p}$ and $\Gamma^\varepsilon_p$ in the same way as was done in Eq.~(\ref{Eqn:OperatorDefinitions}) except that we introduce a `branch-cut' directed along the $y$-direction, as depicted in Fig.~\ref{Fig:BranchCut} and those $U_{x,p}$ that intersect such “branch-cut” acquire an extra $-1$ factor relative to the representation in Eq.~(\ref{Eqn:OperatorDefinitions}), and are given by:
\be\label{Eqn:BoundaryHopping}
U_{x,p} \rightarrow - i \gamma \gamma'.
\ee
Eq.~(\ref{Eqn:Consistency}) then gives $T_x=-1$. Analogous choices are made for other values of $\{T_x, T_y\}$. 

Thus, in summary, $T_{x,y}$ is the operator that determines whether the fermion has anti-periodic boundary conditions along the $x$-, $y$-directions of the torus, and the representations from Eq.~(\ref{Eqn:OperatorDefinitions}) need to be adjusted by adding an appropriate minus sign along a branch-cut of the torus. Clearly there is a freedom in the representation for choosing the precise shape of the branch-cut and other gauges where the vector potential is spread over more bonds are also possible. In Appendix~\ref{Appendix:Duality}, the mapping in Eq.~(\ref{Eqn:OperatorDefinitions}) is constructed more explicitly using a 2D analog of Jordan-Wigner transformation. There the relation of $T_{x,y}$ to boundary conditions (\ref{Eqn:BoundaryHopping}) is also obtained straightforwardly. 

\begin{figure*}
	
	\includegraphics*[width=0.4\linewidth]{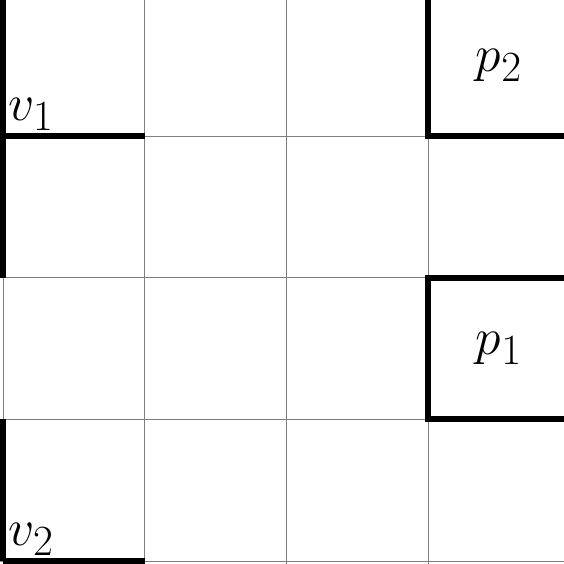}\llap{
		\parbox[b]{13.5cm}{\large (a)\\\rule{0ex}{6.5cm}
	}}~~~~
	\includegraphics*[width=0.4\linewidth]{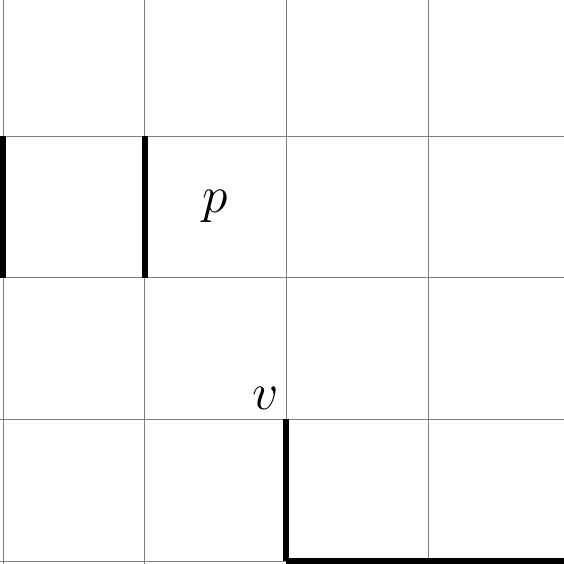}\llap{
		\parbox[b]{13.5cm}{\large (b)\\\rule{0ex}{6.5cm}
	}}%
	\caption{\label{Fig:TCBoundary} (a) $G^e_v$ and $G^m_p$ operators on a boundary in an open lattice. They become three-spin operators on $v_1$, $p_1$. On the lower left vertex $v_2$ and upper right plaquette $p_2$,  $G^e_v$ and $G^m_p$ have only two spins. (b) Creation operators for a single $e$ and $m$ on vertex $v$ and plaquette $p$ respectively, by extending the corresponding $Z$ and $X$ lines from the left and right boundaries. The lattice size is $L_x=L_y=4$.}
	
\end{figure*}

\section{Toric Code and $\mathbb{Z}_2$ charge-flux attachment with open boundaries}\label{Sec:PeriodicLattice}

In this Section we will discuss the detailed implementation of the bosonization construction in lattices with open boundaries. The idea is to first generalize the TC model to a lattice with open boundaries. Provided that the lattice has as many vertices as plaquettes, the $\mathbb{Z}_2$ charge-flux attachment described in Section~\ref{Sec:Review} proceeds then naturally. Open lattices are interesting because they will allow us to explicitly study boundary modes in exactly solvable models that we will describe in Section \ref{Sec:TopologicalSuperconductor}. They are also interesting because the open boundary removes the global parity constraints on the number of non-local $e, m, \varepsilon$ particles. This is because particles appear at the end of string operators but, unlike the torus where the string always has two ends, in open boundaries one can formally view one end of the string to lie outside of the system leaving a single unpaired non-local excitation in its bulk. For related discussion of TC with open boundaries see e.g. Refs.~\onlinecite{Bravyi2002,Kitaev2011}

\subsection{Open boundaries}\label{Sec:OpenLattice}

Our open rectangular lattice is constructed by removing the links along upper and right edges of the rectangular lattice, as shown in Fig.~\ref{Fig:TCBoundary}. The number of links, and consequently of physical local spins, in the lattice is still $2L_xL_y$, and its Hilbert space dimension $2^{2L_xL_y}$. The number of vertices and plaquettes in the lattice is still $L_xL_y$ respectively. The vertex and plaquette operators are defined as:
\be
G^e_v = \prod_{l\in v} X_l,~G^m_{l}=\prod_{l\in p} Z_l,
\ee
where $l$ are the links connected to a given vertex $v$ or surrounding a given plaquette $p$. Notice that the vertex operators, $G^e_v$, acting on the left and bottom edges contain only three links, and the one in the bottom left corner contains only two links, as shown in Fig.~\ref{Fig:TCBoundary}. Similarly, the plaquette operators acting over the top and right edges contain three links and the one in the upper right corner contains 2 links, as shown in Fig.~\ref{Fig:TCBoundary}. However the local algebraic properties of these operators are the same as in those in the usual torus geometry, namely, they are fully commutative among themselves and they have spectrum $\pm1$. However, one important global distinction with the torus is that these operators are completely independent from each other, and in particular they do not satisfy any global parity constraint analogous to that in Eq.~(\ref{Eqn:Constraints1}). We provide a rigorous proof of this in Appendix~\ref{Appendix:TCOperatorIndependence}. As a consequence, the corresponding TC Hamiltonian, given by:
\be\label{Eqn:TCHamiltonian}
H = \Delta_e \sum_v\bigg(\frac{1-G^e_v}{2}\bigg)+\Delta_m \sum_p\bigg(\frac{1-G^m_p}{2}\bigg),
\ee
has a unique ground state and there is a gap, $\min(\Delta_e,\Delta_m)$, to all excitations (assuming $\Delta_{e,m}>0$). This is in agreement with the known property of the ordinary TC topological order, namely that it is not forced to have accompanying gapless boundary modes (see e.g. Ref.~\onlinecite{Levin2013}).

Importantly, in this geometry the $e$ and $m$ particles can be created as isolated particles by a string that extends up the boundary without any accompanying boundary energy cost. In the case of $e$ particles, for example, a string of $Z$ operators can be extended from the location of the $e$ particle towards the right edge or the upper edge, and in the case of the $m$ particles, a string of $X$ operators it can be extended from the desired plaquette towards the bottom or left edge, as depicted in Fig.~\ref{Fig:TCBoundary}. In other words, there are $2L_xL_y$ independent labels associated with $G^e_v, G^m_p$ that can be used to uniquely label the full $2^{2L_xL_y}$-dimensional Hilbert space. Therefore we can view $e$ and $m$ as hard-core bosons without any global parity constraint. If we hold one of these species static, say $e$, by enforcing the commutativity of the Hamiltonian with its local particle number operator, $G^e_v$, then the remaining Hilbert spaces can be exactly mapped into Hilbert spaces of hard-core bosons coupled to static $\pi$-fluxes located at the vertices that contain $e$-particles, without any global parity constraints.

We will now extend $\mathbb{Z}_2$ charge-flux attachment in Ref.~\onlinecite{Chen2018} to open lattices. We begin by describing the modified parity operators that measure the presence of the $\varepsilon$ and $e$ particles. We again view the $e$-particles as residing in the vertices and the $\varepsilon$-particles in the plaquettes. Notice that our lattice has been chosen so that there is a unique plaquette to the north-east of any given vertex, and thus we can follow the same convention of north-east pairing of vertices and plaquettes from the torus defined in Section~\ref{Sec:Review}. The operators measuring the parity of the $e$- and $\varepsilon$-particles are:
\be
\Gamma^e_v = G^e_v\times G^m_{NE(v)},~\Gamma^\varepsilon_p = G^m_p,
\ee
where $NE(v)$ is the plaquette north-east of the vertex $v$. To map onto pure fermionic models we freeze the dynamics of $e$-particles ($\pi$-fluxes) as before, by demanding that the Hamiltonian commutes with every $\Gamma^e_v$ for all vertices $v$. This leads to operators in the bulk which are analogous to those we had in the torus, but forbids certain boundary operators. Namely, we define $U_x$ and $U_y$ in an identical way to how they are defined in Fig.~\ref{Fig:OperatorDefinitions} and Eq.~(\ref{Eqn:PairCreation}).

However, if one of the links making up the $U_{x,y}$ is absent in our new lattice with removed boundaries (see Fig.~\ref{Fig:TCBoundary}), then the corresponding operator $U_{x,y}$ will not commute with some $\Gamma^e_v$ and thus it is not allowed. The remaining allowed operators can be represented exactly as Majorana fermion bilinears as before. Specifically, we introduce two Majorana modes $\gamma, \gamma'$ on every plaquette and we associate the operators in the same way as in Eq.~(\ref{Eqn:OperatorDefinitions}). Such representation from Eq.~(\ref{Eqn:OperatorDefinitions}) would describe the sector $\Gamma^e_v=1$ which has no $e$-particles ($\pi$-fluxes). The sectors with $e$-particles can be represented by introducing strings that connect to the $e$-particles and twisting the sign of the representation of $U_p$ when the fermions hop along such cuts to account for the localized $\pi$-fluxes.~\cite{Oscar2020}

We emphasize that in the current lattice the particle numbers of $\varepsilon$-particles on plaquettes, $(1-\Gamma^\varepsilon_p)/2$, and the particle numbers of the $e$-particle at vertices, $(1-\Gamma^e_v)/2$, form a complete set of labels of all the $2^{2L_xL_y}$ states in the Hilbert space, because there are no global parity constraints on $\varepsilon$ and $e$ in the open lattice, in analogy to the bosonic representation in terms of the parity hard-core bosons $m$ and $e$, discussed at the beginning of this Section. Consequently, we can also create isolated $\varepsilon$-fermions in this geometry by extending the string operators to the boundaries. This allows for a detailed and explicit lattice representation of all operators within any given sector with fixed $\Gamma^e_v$, including the single Majorana mode operator. We note however that the operators with odd fermion parity are necessarily accompanied by non-local strings, whereas the non-local strings disappear from the bilinear operators defined in Eq.~(\ref{Eqn:OperatorDefinitions}), and thus these are the only ones that one must include in physical Hamiltonians or other local operators that are obtained by products of these. Details of the representation of single fermion operator in terms of spin operators in this lattice are presented in Appendix~\ref{Appendix:FermionCreation}.

\begin{figure*}
	
	\includegraphics*[width=0.4\linewidth]{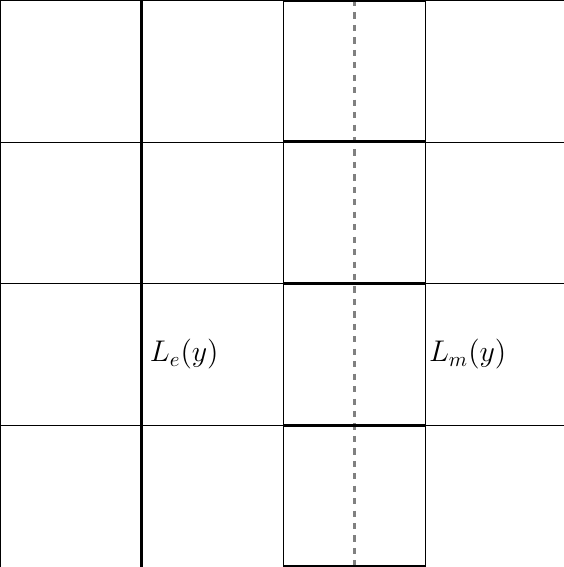}\llap{
		\parbox[b]{13.5cm}{\large (a)\\\rule{0ex}{6.5cm}
	}}~~~~
	\includegraphics*[width=0.4\linewidth]{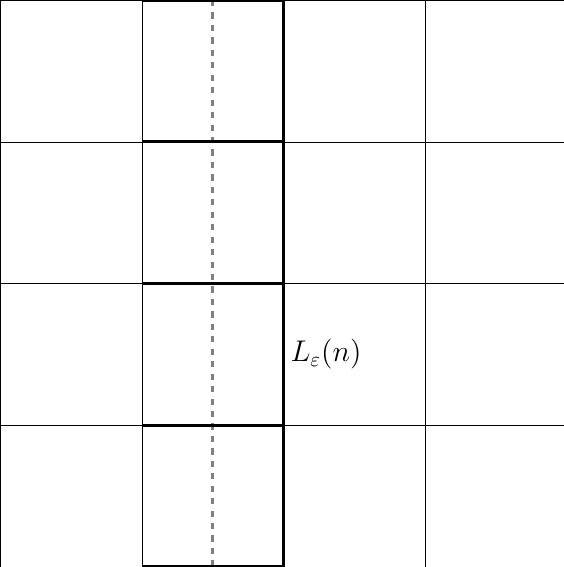}\llap{
		\parbox[b]{13.5cm}{\large (b)\\\rule{0ex}{6.5cm}
	}}%
	\caption{\label{Fig:BoundaryTwistCylinder} (a) The twist of boundary conditions for $e$ and $m$ particles along the $y$-direction in a cylinder. The bold links along paths $L_e(y)$ and $L_m(y)$ are multiplied by $Z$ and $X$ respectively in Eq.~(\ref{Eqn:TCCylinderTwist}). (b) The twist of boundary conditions $T_y$ from Eq.~(\ref{Eqn:t'Hooft1}) for $\varepsilon$ particles along the $y$-direction in a cylinder. $L_\varepsilon(n)$ are along the $n$-th column and $T_y$ along this path satisfies Eq.~(\ref{Eqn:CylinderTwist}).}
	
\end{figure*}

\subsection{Cylindrical Geometry}\label{Sec:CylindricalLattice}

The cylinder geometry has an interesting blend of topological features from the Torus and open lattice geometries. To construct it, we choose the system to be periodic along the $y$-direction and open along the $x$-direction by removing the links in the right edge, as shown in Fig.~\ref{Fig:BoundaryTwistCylinder}.

Operators on the boundary plaquettes with links removed are modified in the same way as the open lattice case. This means $G^e_v$ and $G^m_p$ are three-spin operators on the left and right edges respectively. As in the case of the open lattice, these operators are still completely independent and do not satisfy any global parity constraint, and the $e$ and $m$ particles can still be created as single isolated particles by extending their string towards right and left the open edges of the cylinder respectively. Therefore, the corresponding TC Hamiltonian from Eq.~(\ref{Eqn:TCHamiltonian}) has a unique ground state in the cylinder and a gap to all excitations. Notice that the closed-loop electric and magnetic string operators along the periodic $y$-direction are not independent operators from the local $G^e_v$ and $G^m_p$, but are related by:
\be\label{Eqn:TCCylinderTwist}
\begin{split}
&\prod_{l\in L_m(y)} X_l = \prod_{v\in \text{left of }L_m(y)}G^e_v,\\
&\prod_{l\in L_e(y)} Z_l = \prod_{p\in \text{right of }L_e(y)}G^m_p.
\end{split}
\ee
Here $L_m(y)$ and $L_e(y)$ are closed loops around the periodic $y$-direction associated with transport of $m$ and $e$ particles and the convention for the above relations is depicted in Fig.~\ref{Fig:BoundaryTwistCylinder}.

Now since every vertex has a unique north-east plaquette we can follow the same convention for the $\mathbb{Z}_2$ charge-flux attachment of previous Section, by enforcing that all terms in the Hamiltonian commute with the new $e$-particle parity operator $\Gamma^e_v=G^e_vG^m_{NE(v)}$. This leads to an effective fermionic representation for the various subspaces of the Hilbert space in terms of $\varepsilon$-fermions, whose parity is measured again by $\Gamma^\varepsilon_p=G^m_p$. And we follow the same convention for representation of operators in terms of the two Majorana modes $\gamma, \gamma'$ on every plaquette as the one described in the previous Sections. There are no global constraints on the parity of $\varepsilon$-particles and a single particle creation operator can be defined. But it always involves a non-local loop operator, and therefore can be discarded from appearing in physical Hamiltonians, which will only contain again operators within the fermion parity even sub-algebra and thus can be completely generated by from the local spin operators $\Gamma^\varepsilon_p,U_{x,p},U_{y,p}$.

One particularly amusing aspect of the cylinder geometry is that, even though there are no global parity constraints on the $\varepsilon$ particles, it is still possible to twist boundary conditions along the periodic $y$-direction. At first glance one might think that this will induce a mismatch between the size of the dual fermionic Hilbert space and that of the underlying spin Hilbert space, since  the locations of $\varepsilon$-fermions and $\pi$-fluxes are enough to label all the states in the physical Hilbert and exhaust its dimensionality, and thus one might think the extra twist of boundary conditions along $y$-direction will double the size of the dual fermionic Hilbert space relative to the underlying spin space. There is however a non-trivial constraint between the local $e$ and $\varepsilon$ parity operators in the fermionic operators and the operator that transports fermions over a closed loop around the $y$-periodic direction of the cylinder, $T_y$. Namely by adopting the same definition we had in the torus in Eq.~(\ref{Eqn:t'Hooft1}) for the operator $T_y$ that performs transport over the periodic direction, we encounter that this operator satisfies the following constraint with products of local parity operators of $e$ and $\varepsilon$ particles : 
\be\label{Eqn:CylinderTwist}
T_y = -\bigg(\prod_{\text{left of}~L_\varepsilon(n)} \Gamma^e_v\bigg)  \bigg(\prod_{p\in \text{lattice}}\Gamma^\varepsilon_p \bigg),
\ee
where $L_\varepsilon(n)$ is a vertical closed loop around the periodic $y$-direction at the n-th column of the lattice. The schematic of the definition of these operators is depicted in Fig.~\ref{Fig:BoundaryTwistCylinder}. The first product of $\Gamma^e_v$ operators can be understood intuitively by noting that it measures the extra induced twist of boundary conditions by the presence of static $e$ particles ($\pi$-fluxes), within the convention that $e$-particles are added from the right open edge of the cylinder, and that each one induces a $-1$ twist of the amplitude of the hopping in the vertical $y$-direction, as depicted in Fig.~\ref{Fig:TCBoundary}. The second product of $\Gamma^\varepsilon_p$ is very interesting as it implies that the the boundary conditions along the $y$-direction are not independent of the global parity of the fermions. In particular in the case of no static $e$-particles ($\Gamma^e_v=1$ for all $v$), the constraint implies that for a total odd (even) number of $\varepsilon$-fermions in the cylinder one must necessarily choose periodic (anti-periodic) boundary conditions along its $y$-direction. In other words, the dual Hilbert spaces with e.g. periodic $y$-boundary conditions and an even number of fermions must be discarded as un-physical. 

\begin{figure}
	
	\includegraphics*[width=0.8\linewidth]{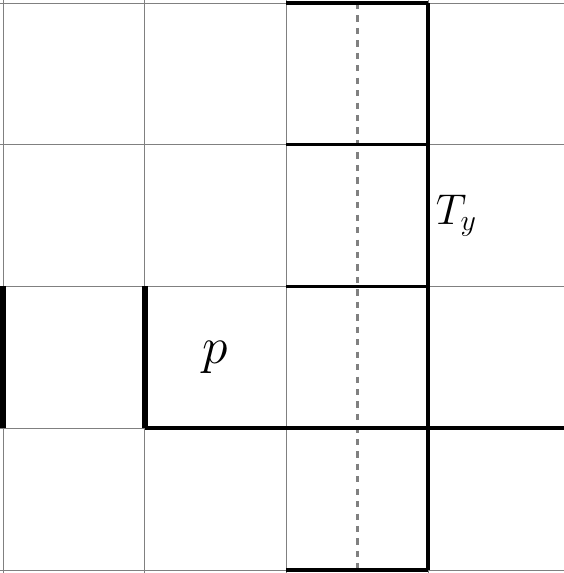}
	\caption{\label{Fig:FermionCreationCylinder} String operator that changes $\varepsilon$-parity at plaquette $p$. The operator intersects $T_y$ and changes the vertical boundary condition for $\varepsilon$-fermions. This is due to the dependence of $\varepsilon$ particle number and vertical twist $T_y$ in Eq.~(\ref{Eqn:CylinderTwist}).}
	
\end{figure}

There is a simple intuitive picture behind this amusing constraint, which is illustrated in Fig.~\ref{Fig:FermionCreationCylinder}. From Fig.~\ref{Fig:FermionCreationCylinder} one can see that this constraint arises from the fact that operators that raise the $\varepsilon$-fermion number by one without adding $e$-particles must have electric and magnetic strings extending to opposite open edges of the cylinder, and therefore they intersect $T_y$ an odd number of times leading to these operators to anti-commute, and thus to the property that the boundary conditions and the global fermion parity cannot be changed independently but must obey the constraint in Eq.~(\ref{Eqn:CylinderTwist}). This point is further discussed in Appendix~\ref{Appendix:FermionCreation}. The above discussion implies that in order to properly dualize the subspaces with static $e$-particles (commutativity with every $\Gamma^e_v$) as ordinary fermionic models of $\varepsilon$-particles, one must impose a global fermion parity conservation, namely that the Hamiltonian commutes with $\Pi_{p\in \text{lattice}} \Gamma^\varepsilon_p$, in order to have a definite  fermionic boundary condition along the periodic $y$-direction.

\section{Topological Superconductors with Translational Symmetry}\label{Sec:TopologicalSuperconductor}

The exact fermionic representations of spin Hamiltonians in terms of fermionic models described in previous Sections provides a boundless tool to build new phases of matter on top of the Toric Code vacuum. Naturally a simple class of phases is that in which $\varepsilon$-fermions have an effective non-interacting fermion bilinear Hamiltonian. The only `unbreakable symmetry' that these $\varepsilon$-fermions are required to have is their global parity. Therefore the natural free-fermion states that one is lead to consider are those described by Bogoliubov-De-Gennes (BdG)-type Hamiltonians. In two dimensions and in the absence of any symmetry, these are Hamiltonians belonging to class D and in the topological classification of free particle systems are labeled by the integer spectral Chern number, $C\in \mathbb{Z}$, which counts the number of right-moving minus the number of left-moving Majorana modes at the edge.~\cite{Ryu2010} The $\mathbb{Z}_2$ topologically ordered states that one would construct on top of the Toric Code vacuum by having the $\varepsilon$-fermions form a topological superconductor state with Chern number $C$ were those considered by Kitaev in his seminal paper Ref.~\onlinecite{Kitaev2006}, where he demonstrated that the bulk topological properties of the anyons in such phases, as encoded in the data of their fusion modular tensor category, only depend on $C \mod 16$. In spite of this, any two states with different $C$ can still be regarded as topologically distinct phases since they cannot be connected adiabatically while preserving their bulk gap.

In the present study we would like to extend these considerations to the case in which the topological order is enriched only by the discrete lattice translational symmetry. We will restrict to cases in which the $e$-particles ($\pi$-fluxes) are absent, which means that we will only consider the phases in which the translational symmetry is implemented non-projectively on the $\varepsilon$-fermions. In the perspective of the projective symmetry group of Refs.~\onlinecite{Wen2002,Wen20021}, these correspond to states where the $\varepsilon$-fermions experience zero flux per unit cell. Another set of translational invariant states are those in which there is one $e$-particle ($\pi$-flux) in every vertex, which can be studied by similar methods to those we develop, but we will not consider this case here. However, as we will see in Section~\ref{Sec:WeakSymmetryBreaking}, some of the phases that we will consider still feature a non-trivial projective representation of the translational symmetry of $e$-particles. Therefore, we are naturally led to consider the symmetry protected topological phases of free fermions in Class D enriched by translational symmetry. The remainder of this Section is essentially a review of results in the literature of classification of BdG Hamiltonians with particular emphasis on the aspects that are relevant for our analysis. We note in passing that even though our analysis is restricted to only BdG Hamiltonians with discrete translational symmetries, it can be naturally extended to other symmetries, which is naturally aided by recent progress on completing the full classification of crystalline topological BdG Hamiltonians.~\cite{Ono2020,Geier2020,Schindler2020,Ono2021} 

\subsection{$\mathbb{Z}\times(\mathbb{Z}_2)^3$ classification of translationally invariant 2D BdG Hamiltonians}
We assume the fermion bilinear Hamiltonian has an ordinary commutative discrete translational symmetry group with generators $\{t_x,t_y\}$. This requires that fermion pairing terms respect translational symmetry and therefore Cooper pairs carry zero momentum. We can therefore label BdG fermion eigenmodes by crystal momenta $(k_x, k_y)$. In crystal momentum basis, the BdG Hamiltonian pairs states of momenta $\bm{k}$ and $-\bm{k}$. There are four special momenta residing at the center and corners of the Brillouin zone that satisfy $\bm{k}=-\bm{k}~~\text{mod}~ 2\pi$, namely $\{(0,0),(0,\pi),(\pi,0),(\pi,\pi)\}$. They are special because the fermion modes at these momenta are `paired with themselves'. Therefore, for these points the BdG Hamiltonian can be viewed effectively as a 0D single site Hamiltonian. 0D BdG Hamiltonians (class D) are in turn classified by a $\zeta\in \mathbb{Z}_2$ index,~\cite{Ryu2010} which simply measures the parity of the fermion number operator $(N_F)$ at the site, $\zeta=N_F~~\text{mod}~ 2$. Namely, $\zeta=0$ corresponds to states with an even number of fermions on the site, which are adiabatically connected to the trivial empty vacuum with no fermions, and $\zeta=1$ corresponds to states with odd fermions on the site, which are connected adiabatically to the state with only one fermion. As a consequence, topological superconductors with translational symmetry in 2D have four topologically invariant $\mathbb{Z}_2$ indices (also referred to as Pfaffian indicators),~\cite{Geier2020} which measure the fermion number parity at the $4$ special momenta in the Brillouin zone.~\cite{Wen2009,Wen2010,Geier2020} We will represent these $4$ parity indices with a $2\times2$ matrix, $\zeta_{ij}$, where the indices $i,j$ denote the special momenta $\bm{k}_{ij}$, arranged as follows:
\be\label{Eqn:MomentaMatrix}
\begin{pmatrix} (0,0) & (0,\pi) \\ (\pi,0)&  (\pi,\pi)\end{pmatrix}\equiv\begin{pmatrix} \bm{k}_{11} &\bm{k}_{12}\\\bm{k}_{21}&  \bm{k}_{22}\end{pmatrix},~\zeta_{ij} = \zeta (\bm{k}_{ij}).
\ee
These topological parity indices are not all independent from the spectral Chern number, $C\in \mathbb{Z}$, but satisfy the following constraint:~\cite{Sato2010,Sato12010}
\be\label{Eqn:ChernNumberConstraint}
(-1)^C = \prod_{i,j=1}^2 (-1)^{\zeta(\bm{k}_{ij})}
\ee
Therefore, once the Chern number $C$ is specified, only three of the parity labels are independent, and we have a $\mathbb{Z}\times (\mathbb{Z}_2)^3$ classification of translationally invariant topological superconductors in 2D.

\begin{figure}
	\includegraphics*[width=0.5\linewidth]{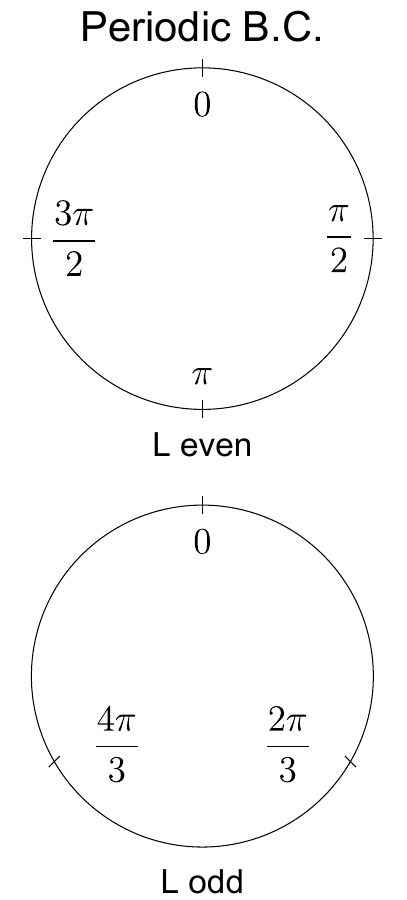}~	
	\includegraphics*[width=0.5\linewidth]{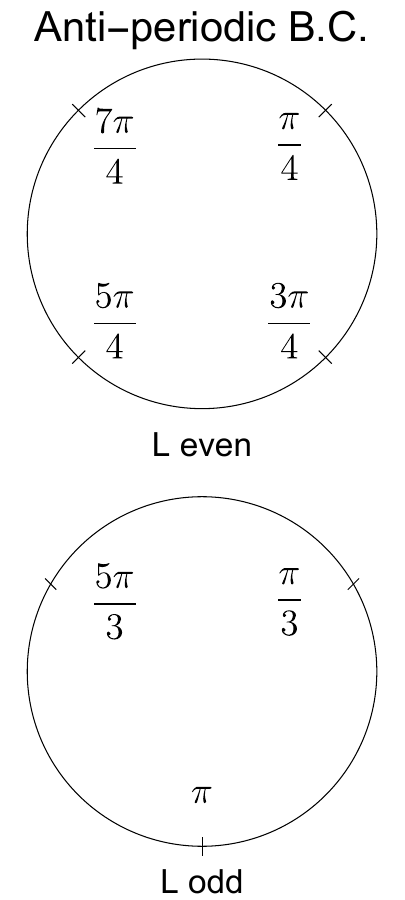}
	\caption{\label{Fig:MomentumQuantisation} Visual representation of quantisation of momenta for periodic and anti-periodic boundary conditions and given lattice size $L$. We show here odd $L=3$ and even $L=4$. }
\end{figure}

To illustrate this more concretely, let us consider a BdG Hamiltonian with a single complex fermion mode, $a_{\bm{R}}$, on every unit cell (spinless fermions with a single site per unit cell) labeled by the vector $\bm{R}$ in the Bravais lattice. These systems are sufficient to realize representatives of all the topologically non-trivial phases and the exactly solvable models that we will discuss in Section~\ref{Sec:Model} are of this kind. In crystal momentum basis $a_{\bm{k}}^\dagger=N^{-1/2}\sum_{\bm{R}} \exp(-i\bm{k}.\bm{R}) a_{\bm{R}}^\dagger$, the BdG Hamiltonian has the form:
\be\label{Eqn:BCSHamiltonian}
H =\sum_{\bm{k}} \Psi_{\bm{k}}^\dagger \begin{pmatrix}\varepsilon(\bm{k})& \Delta(\bm{k})\\\Delta^*(\bm{k})&-\varepsilon(-\bm{k}) \end{pmatrix}\Psi_{\bm{k}},~\Psi_{\bm{k}}=\begin{pmatrix} a_{\bm{k}}\\ a_{-\bm{k}}^\dagger\end{pmatrix}.
\ee
The pairing function is antisymmetric $\Delta(\bm{k})=-\Delta(-\bm{k})$, and therefore at the special momenta satisfying $\bm{k}_{ij}=-\bm{k}_{ij}$, the BdG Hamiltonian is diagonal and the sign of $\varepsilon(\bm{k}_{ij})$ determines the topological parity index $\zeta_{ij}$. Namely, the complex fermion mode at $\bm{k}_{ij}$ is occupied if $\varepsilon(\bm{k}_{ij})<0$ and empty if $\varepsilon(\bm{k}_{ij})>0$. The topological index $\zeta_{ij}$ is therefore simply given by the zero temperature Fermi-Dirac occupation function at such momenta,~\cite{Wen2009,Wen2010} which explicitly reads as:
\be\label{Eqn:TopologicalIndex}
\zeta_{ij} = 1 - \Theta[\varepsilon(\bm{k}_{ij})],
\ee
These $\zeta_{ij}$ parity indices determine also the global fermion number parity of the ground state when placed on a finite size torus,~\cite{Wen2009,Wen2010} in a way that generalizes the classic result of Read and Green on 2D topological paired states.~\cite{Read2000} To see this we consider a finite torus with a number of $L_{x,y} \in \mathbb{Z}$ Bravais unit cells along the $x$-, $y$-directions, whose crystal momenta belong to a discrete lattice:
\be
(k_x,k_y)=2\pi \bigg[\frac{n_x+\Phi_x/(2\pi)}{L_x},\frac{n_y+\Phi_y/(2\pi)}{L_y} \bigg],~ n_{x,y}\in \mathbb{Z}.
\ee
Here we imagine that the system can have periodic or anti-periodic boundary conditions along the two directions of the torus leading to twists of boundary conditions labeled by $\Phi_{x,y}\in \{0,\pi\}$. Crucially, some of the special crystal momenta might not be allowed in a given finite size torus depending on the parity of the total number of unit cells $L_{x,y}~\text{mod}~2$ and the boundary condition twist. This is illustrated in Fig.~\ref{Fig:MomentumQuantisation} where crystal momenta are depicted as discrete angles in a circle. It is useful to construct a matrix, $A(\bm{k}_{ij})$, of `allowed' momenta, namely a function which equals $1$ when a special crystal momentum point $\bm{k}_{ij}$ is allowed and 0 when it is not in a given system:
\be\label{Eqn:AFunction}
\begin{split}
A(\bm{k}_{ij}) &= X(k_{x,ij}) Y(k_{y,ij}),\\
X(k_x) &=\bigg( 1-\frac{\Phi_x}{\pi}\bigg) + (-1)^{L_x}\frac{k_x}{\pi}\bigg[ 1-\frac{\Phi_x}{\pi} - (L_x ~\text{mod}~ 2) \bigg],
\end{split}
\ee
where $Y(k_y)$ is obtained from $X(k_x)$ by exchanging all of the `$x$' by `$y$' labels in the expression above. Therefore the total fermion particle number parity of a ground state in a finite torus can be simply obtained by adding the topological parity index $\zeta_{ij}$ that counts the parity of fermion occupation at the special momentum $\bm{k}_{ij}$, weighed by the function $A(\bm{k}_{ij})$ that equals $1$ if the corresponding special momentum is allowed and $0$ otherwise and it is explicitly given by the following formula:
\be\label{Eqn:FermionParityMatrix}
N_f ~\text{mod}~2= \bigg(\sum_{i,j=1}^{2} A(\bm{k}_{ij}) \zeta_{ij} \bigg) ~\text{mod}~ 2 = \Tr (A^T \zeta) ~\text{mod}~ 2.
\ee
In the second equality, $\zeta_{ij}$ and $A(\bm{k}_{ij})$ are viewed as matrices with momenta index $i, j$ arranged as described in Eq.~(\ref{Eqn:MomentaMatrix}). Table~\ref{Table:AFunction} lists the $A$ matrices for the various twist and parities of the number of lattice sites. This matrix notation should simplify the bookkeeping of determining when a BdG topological phase has an odd number of fermions in a finite torus, by simply taking the sum of the component-by-component product of the $\zeta$ and $A$ matrices and determining if it is even or odd from Eq.~(\ref{Eqn:FermionParityMatrix}).
\begin{table}
		\centering
	\begin{tabular}{|c|c|c|c|c|}
		\hline
		~   & (P-P) & (AP-P) & (P-AP) & (AP-AP) \\ \hline
		(e-e)& $\begin{pmatrix}1&1\\1&1	\end{pmatrix}$& $\begin{pmatrix}0&0\\0&0\end{pmatrix}$& $\begin{pmatrix}0&0\\0&0\end{pmatrix}$& $\begin{pmatrix}0&0\\0&0\end{pmatrix}$\\\hline
			(e-o) & $\begin{pmatrix}1&0\\1&0\end{pmatrix}$&  $\begin{pmatrix}0&0\\0&0\end{pmatrix}$&  $\begin{pmatrix}0&1\\0&1\end{pmatrix}$& $\begin{pmatrix}0&0\\0&0\end{pmatrix}$ \\\hline
			(o-e) & $\begin{pmatrix}1&1\\0&0\end{pmatrix}$& $\begin{pmatrix}0&0\\1&1\end{pmatrix}$& $\begin{pmatrix}0&0\\0&0\end{pmatrix}$&$\begin{pmatrix}0&0\\0&0\end{pmatrix}$ \\\hline
			(o-o) &$\begin{pmatrix}1&0\\0&0\end{pmatrix}$& $\begin{pmatrix}0&0\\1&0\end{pmatrix}$& $\begin{pmatrix}0&1\\0&0\end{pmatrix}$&$\begin{pmatrix}0&0\\0&1\end{pmatrix}$\\ \hline  
		\end{tabular}
			\caption{ $A$ matrix for all lattice size and boundary conditions}
\label{Table:AFunction} 
\end{table}

\subsection{Lower dimensional stacking and bulk-boundary correspondence}

Let us now discuss the real space picture of this finer topological classification of 2D translationally invariant BdG Hamiltonians and its manifestations in terms of gapless boundary modes in open lattices. Interestingly, some but not all of the states with non-trivial $\mathbb{Z}\times (\mathbb{Z}_2)^3$ labels have boundary gapless modes. These parity labels are indeed an example of `weak topological' indices, in an analogous sense to those in time-reversal-invariant topological insulators,~\cite{Moore2007,Fu2007} namely, they characterize stacking patterns of lower dimensional topological phases,~\cite{Sato2010,Sato12010,Geier2020} and have therefore a very transparent real space interpretation. In order to understand such real space interpretation of these indices in 2D, it is useful to understand the classification of lower dimensional BdG Hamiltonians with translational symmetry, which we shall review next.

Topological superconductors without symmetry (class D) in 0D and 1D both have $\mathbb{Z}_2$ topological classifications.~\cite{Ryu2010} In 0D, the state with trivial $\zeta=0$ $\mathbb{Z}_2$ index, has an even number of fermions in the site, while the non-trivial state, $\zeta=1$, has an odd number of fermions in the site. In 1D, the trivial state with $K=0$ $\mathbb{Z}_2$ index is connected adiabatically to the trivial vacuum with zero fermions per site, while the non-trivial state with $K=1$ $\mathbb{Z}_2$ index has an odd number of unpaired Majorana modes at each end of the wire, and its classic realization is the Kitaev wire model.~\cite{Kitaev2001} With translational symmetry in 1D there appear two additional weak $\mathbb{Z}_2$ invariants, $\zeta(k_i)\in \{ 0,1\}$, measuring the fermion parity at the two special momenta $k_i\in \{ 0,\pi\}$ analogously to the 2D case discussed above. These weak parity invariants are constrained by the strong 1D topological index $K$, as follows:~\cite{Geier2020}
\be
(-1)^K = \prod_{i=1}^2 (-1)^{\zeta_{k_i}}.
\ee
Therefore, 1D BdG superconductors (class D) with lattice translations, can be fully classified by two independent $\mathbb{Z}_2\times\mathbb{Z}_2$ labels $(\zeta_0,\zeta_\pi),$ and there is, therefore, a total of $4$ topologically distinct phases. The two states $(0,0)$ and $(1,1)$ with trivial strong label $(K=0)$ are adiabatically connected to the `stacks' of 0D dimensional phases, and are therefore  `weak' topological states. Specifically, the trivial $(0,0)$ phase is adiabatically connected to the trivial vacuum with no fermions per site, while the $(1,1)$ phase is adiabatically connected to the stack of 0D sites with one fermion per site. This can be seen simply by noting that an insulator with a fully occupied band with one fermion per site would have occupied both special 1D momenta $k_i\in \{ 0,\pi\}$. Therefore these states are `Atomic Insulators' (AI),~\cite{Geier2020} and clearly have no dangling gapless edge Majorana modes. We note that, because of the above, in the classification convention of Ref.~\onlinecite{Geier2020}, the state $(1,1)$ is viewed as a `trivial' state because it has a trivial `atomic insulator limit'. However, for our purposes it is important to keep track of this phase as a non-trivial topologically distinct phase from $(0,0)$ because they cannot be connected adiabatically without closing the bulk gap. In fact this distinction is robust beyond non-interacting BdG Hamiltonians, because the $(1,1)$ ground state has a global odd number of fermions in 1D chains with an odd number of sites regardless of twist of boundary conditions and an even number of fermions in lattices with an even number of sites, in sharp contrast to the $(0,0)$ state which always has even number of fermions regardless of twist and parity of the number of lattice sites. This will be particularly important in our case because states with an odd number of fermions must be discarded as unphysical when the fermions are emergent and are microscopically forced to be created only in pairs from a topologically ordered ground state in the torus, as it is the case of the $\varepsilon$-fermions previously discussed in Section~\ref{Sec:Review}. This is in fact the underlying cause of the anomalous ground state degeneracy in the torus of certain $\mathbb{Z}_2$ topologically ordered states discussed in Refs.~\onlinecite{Wen2003,Kou2008,Wen2009,Wen2010,Cho2012}, which we will review in the forthcoming Sections.

The states with labels $(1,0)$ and $(0,1)$ are strong 1D topological superconductors ($K=1$ Kitaev-wire-type states) which are obtained from the trivial state $(0,0)$ via a phase transition by closing the gap either at $k=0$ or $k=\pi$ respectively. They both feature an odd number of dangling Majorana modes at each edge, and can be distinguished by their global fermion parity in finite periodic chains with $L\in \mathbb{Z}$ sites subjected to periodic ($\Phi=0$) and anti-periodic ($\Phi=\pi$) boundary conditions. Specifically, the following formula, which is the 1D analogue of Eq.~(\ref{Eqn:FermionParityMatrix}), gives the number of fermions in a periodic chain:
\be
N_f ~\text{mod}~2= \bigg(\sum_{i=1}^{2} A(k_i) \zeta(k_i) \bigg) ~\text{mod}~ 2 = \bm{A}.\bm{\zeta}.
\ee
$A(k_i) = X(k_i)$ and the function $X(k_i)$ is the same as in Eq.~(\ref{Eqn:AFunction}). This formula predicts that the state $(1,0)$ will have an odd (even) number of fermions in its ground state under periodic (anti-periodic) boundary conditions regardless of the number $L$ of lattice sites. On the other hand $(0,1)$ will have an odd number of fermions for chains with $L$ even and periodic boundary conditions and $L$ odd and anti-periodic boundary conditions, and otherwise it will have an even number of fermions.

Armed with the above results in $0D$ and $1D$, we are now in a position to understand the real space picture of the $\mathbb{Z}\times (\mathbb{Z}_2)^3$ topological classification of BdG superconducting phases with translational symmetry in 2D. First notice that if we construct a 2D BdG systems out of stacks of decoupled 1D wires which extend along the $x$- ($y$-)direction, then the parity index matrix $\zeta_{ij}$ will be independent of its $i$-component ($j$-component). This implies that the following phases will be adiabatically connected to 0D atomic insulators insulators ($\text{AI}_i$) with an even ($i=0$) and odd ($i=1$) number of fermions per site respectively:
\be\label{Eqn:Phases1}
\text{AI}_0:~\zeta_{ij}=\begin{pmatrix}0&0\\0&0\end{pmatrix};~\text{AI}_1:~\zeta_{ij}=\begin{pmatrix}1&1\\1&1\end{pmatrix}.
\ee
Neither of the atomic insulators, $\text{AI}_i$, has dangling Majorana modes at the boundaries. $\text{AI}_0$ has always an even number of fermions in its ground state regardless of the parity of the torus size or the twist of boundary conditions, whereas $\text{AI}_1$ has a fermion parity that equals the parity of the number of sites in the lattice $L_xL_y ~\text{mod}~2$ independent of the twist of boundary conditions. Similarly the following phases are adiabatically connected to decoupled stacks of Kitaev-wires $(\text{KW}_{\alpha,\zeta})$ aligned along the $\alpha$-directions ($\alpha\in\{x,y\}$) and with a 1D parity index $\zeta$ at $k=\pi$ ($\zeta\in\{0,1\}$):
\begin{subequations}
\be\label{Eqn:Phases2} 
\text{KW}_{x,0}: ~\zeta_{ij}=\begin{pmatrix}1&1\\0&0\end{pmatrix};~\text{KW}_{x,1}: ~\zeta_{ij}=\begin{pmatrix}0&0\\1&1\end{pmatrix}; 
\ee
\be\label{Eqn:Phases3}
\text{KW}_{y,0}: ~\zeta_{ij}=\begin{pmatrix}1&0\\1&0\end{pmatrix};~\text{KW}_{y,1}: ~\zeta_{ij}=\begin{pmatrix}0&1\\0&1\end{pmatrix}.
\ee
\end{subequations}

When placed on a lattice with open boundaries, $\text{KW}_{\alpha,\zeta}$ phases will have an odd number of dangling Majorana modes per exposed unit cell along the open boundaries that are orthogonal to the $\alpha$-direction and an even number of Majorana modes per exposed unit cell for boundaries parallel to the $\alpha$-direction of the wires, provided the translational symmetry along the boundary is preserved. 

There are two other weak topological phases that are adiabatically connected to decoupled 1D Kitaev wires, and are those in which the $\zeta_{ij}$ parity index depends only on the sum of $i+j~\text{mod}~ 2$. These can be viewed as decoupled Kitaev wires that are oriented along the diagonal direction, namely, the fermion modes in a unit cell labeled by coordinates $(R_x,R_y)$ only couple to fermions in the unit cells given by $(Rx+n,Ry+n)$, with $n\in\mathbb{Z}$. Because of this, we will denote these `diagonal' Kitaev-wire phases by $\text{KW}_{x+y,\zeta}$ where $\zeta \in \{0,1\}$ is the 1D parity index of the wires, and they have topological 2D parity indices given by:
\be\label{Eqn:Phases4}
\text{KW}_{x+y,0}: ~\zeta_{ij}=\begin{pmatrix}1&0\\0&1\end{pmatrix};~\text{KW}_{x+y,1}: ~\zeta_{ij}=\begin{pmatrix}0&1\\1&0\end{pmatrix}.
\ee
When placed on a lattice with open boundaries, the $\text{KW}_{x+y,\zeta}$ phases will have an odd number of dangling Majorana modes per exposed unit cell in all of the boundaries for which the boundary translational symmetry is preserved. The phases in Eqs.~(\ref{Eqn:Phases1})-(\ref{Eqn:Phases4}) exhaust all of the 2D `weak' topological phases that are adiabatically connected to stacks of lower dimensional topological phases. In particular, notice that other `slopes' for stacking of wires do not lead to new topological phases. For example, if we stack Kitaev wires with a `slope' $(q_x,q_y), q_{x,y}\in \mathbb{Z}$, by coupling fermion modes $\gamma$ at the unit cell $(R_x,R_y)$ only with fermion modes $\gamma'$ in the unit cell $(R_x+q_xn,Ry+q_yn)$, with $n\in\mathbb{Z}$, one can show that this state will be topologically equivalent to a state with a different slope $(q_x',q_y')$ provided that $q_{x,y}'=q_{x,y}~\text{mod}~2$. This follows from the fact that these two phases have $\varepsilon(\bm{k})\propto \cos (q_x k_x + q_y k_y)$ and $\varepsilon(\bm{k})\propto \cos (q_x' k_x + q_y' k_y)$, and they have the same topological indices given by Eq.~(\ref{Eqn:TopologicalIndex}) evaluated at $k_{x,y} = 0,\pi$ due to $2\pi$-periodicity of the cosine function. Therefore we see that the AI$_\zeta$,KW$_{x,\zeta}$,KW$_{y,\zeta}$,KW$_{x+y,\zeta}$ phases, which respectively have slopes $(0,0),(1,0),(0,1),(1,1)$, cover all the possible slopes of wire stacking modulo $2$.

The weak topological superconducting phases form a modular additive group, where the physical interpretation of addition is aligning the phases `on top of each other', as in a bilayer system while preserving the translational symmetry. The topological parity matrices, $\zeta_{ij}$, of a decoupled bilayer is the sum of the topological parity matrices of each layer modulo $2$. Because of this we can specify a `complete basis' of phases out of which all other can be obtained by layer addition. This basis would only have 3 phases, which we could choose for example to be KW$_{x,0}$, KW$_{y,0}$,AI$_1$, and the three $\mathbb{Z}_2$-valued coefficients ($0$ and $1$) that specify any other phase in this basis can be taken as the $(\mathbb{Z}_2)^3$ topological labels in the $\mathbb{Z}\times (\mathbb{Z}_2)^3$ classification. Then to complete the basis to generate all of the possible 2D BdG superconducting phases by layer addition, we simply need to specify two non-trivial states with non-zero Chern numbers $C=\pm1$, which we can choose to be the simplest chiral topological topological superconductors, denoted by $\chi_C$, and describe them by a parity matrix:
\be\label{Eqn:ChernNumber2}
\chi_C = \begin{pmatrix}1&0\\0&0\end{pmatrix}.
\ee
These $\chi_C$ topological superconductor has spectral Chern number $C=\pm1$. They can be obtained from the trivial vacuum, AI$_0$, by closing the gap at the special momenta $(0,0)$ and they are a lattice version of the celebrated $p\pm ip$ spinless superconductor described by Read and Green,~\cite{Read2000} with a chiral gapless Majorana boundary mode, and an odd number of fermions in the torus for periodic boundary conditions along $x$- and $y$-directions, and even number otherwise. Therefore $(\chi_C,\text{KW}_{x,0},\text{KW}_{y,0}\text{AI}_1)$ form a complete basis for layer addition for all topological BdG states with translation in 2D, and we can specify any state by a unique vector $(C,\zeta_{K_x},\zeta_{K_y},\zeta_{\text{AI}})\in (\mathbb{Z} ,\mathbb{Z}_2,\mathbb{Z}_2,\mathbb{Z}_2)$.

\subsection{Robustness of $\mathbb{Z}\times (\mathbb{Z}_2)^3$ classification against interactions and disorder}
Our discussion of the $\mathbb{Z}\times (\mathbb{Z}_2)^3$ classification 2D translational invariant topological superconductors has so far been restricted to non-interacting fermion bilinear Hamiltonians, and therefore, a natural question is whether this classification is stable against fermion interactions. In fact, it is known that certain symmetry protected topological superconducting phases are not stable against interactions, such as 1D superconductors with $T^2=+1$ time-reversal (1D BDI class), whose non-interacting $\mathbb{Z}$ classification collapses down to $\mathbb{Z}_8$ under interactions,~\cite{Fidkowski2010,Turner2011,Fidkowski2011,You2014} as well as other examples.~\cite{Ryu2012,Qi2013,Yao2013,Wang2014,Gu2014,Metlitski2014,Lu2016} There is however a simple argument that indicates the $\mathbb{Z}\times (\mathbb{Z}_2)^3$ classification 2D topological superconductors is fully stable against interactions. First, the spectral Chern number $C$ is expected to be stable against interactions. Second, we can provide an alternative definition of the topological parity matrix at special momenta $\zeta_{ij}$ from Eq.~(\ref{Eqn:TopologicalIndex}), in terms of many-body properties without reference to the single particle BdG spectrum. This can be done by noting from Table~\ref{Table:AFunction} that when the system is placed in a torus in which both $L_x$ and $L_y$ are odd, the topological parity index $\zeta_{ij}$ can be defined as the parity of the many fermion ground state, $N_f ~\text{mod}~ 2$, under twists of boundary conditions $N_f (\Phi_x,\Phi_y) ~\text{mod}~ 2$ as follows:
\be\label{Eqn:TopologicalIndex1}
\zeta_{ij} = \begin{pmatrix} N_f(0,0)&N_f(0,\pi)\\N_f(\pi,0)&N_f(\pi,\pi) \end{pmatrix}~\text{mod}~ 2,~L_{x,y}~\text{odd}.
\ee
Since the many-body fermion parity of the ground state will not change by adding interactions, unless a bulk-gap closing phase transition is induced, the topological parity matrix $\zeta_{ij}$ will remain quantized to have $\{0,1\}$ entries and the $\mathbb{Z}\times (\mathbb{Z}_2)^3$ classification of translational invariant superconductors is expected to remain stable upon adding fermion interactions.

The above re-casting of the topological parity matrix also indicates that the $\mathbb{Z} \times (\mathbb{Z}_2)^3$ classification of translational invariant superconductors is stable in the presence of self-averaging disorder that respects translational symmetry. To see this, we appeal again to the fact that disorder is not expected to change the many-body fermion parity of a gapped state unless a bulk phase transition occurs. This is an important point because the label of these states as `weak' topological phases might create the wrong impression that the states would be delicate or fragile. This robustness of `weak' topological labels against disorder has been emphasized previously in the case of time-reversal-invariant weak topological insulators,~\cite{Mong2012,Ringel2012} and topological superconductors with other symmetries.~\cite{Morimoto2014}

\section{Translationally symmetric $\mathbb{Z}_2$ topologically ordered states}\label{Sec:WeakSymmetry}

\subsection{Anomalous GSD In Tori}

As we have seen 2D translationally invariant topological superconductors can have ground states with an odd fermion number in the torus. As first identified in Refs.~\onlinecite{Wen2003,Kou2008,Wen2009,Wen2010,Cho2012}, when such paired fermions are the $\varepsilon$-fermions that emerge in a $\mathbb{Z}_2$ topologically ordered state, where the periodic and anti-periodic boundary conditions are realized dynamically by the Hamiltonian, this leads to an `anomaly' in the number of degenerate topological ground states in the torus. Specifically, as discussed in Section~\ref{Sec:Review}, only states with a global even number of fermions are physical and states with an odd number of fermions must be discarded. Therefore this leads to the following formula for the ground state degeneracy of a $\mathbb{Z}_2$ topologically ordered state where fermions form a translationally invariant paired state of the kind described in Section~\ref{Sec:TopologicalSuperconductor}:
\be\label{Eqn:AnomalousGSD}
\text{GSD} = 4 - \bigg[\sum_{\Phi_x,\Phi_y}\bigg(\Tr A^T\zeta\bigg) ~\text{mod}~ 2\bigg].
\ee
Here the sum is over the twist of BCs, $\Phi_{x,y}\in \{0,\pi\}$, for a phase described by a topological parity matrix $\zeta$ and for a torus with a given number of $L_{x,y}$ unit cells along $x$- and $y$-directions. The $A$ matrices are given by Eq.~(\ref{Eqn:AFunction}) and are tabulated in Table~\ref{Table:AFunction}. Notice that the difference of GSD between two states with different $\zeta$, can in some cases be understood as a manifestation of different bulk topological order but in some others it cannot. For example, as shown by Kitaev,~\cite{Kitaev2006} the bulk topological order of the superconductor depends on the spectral Chern number $C ~\text{mod}~ 16$, and states with even $C$ are expected to have a four-fold GSD, while $C$ odd are expected to have three-fold GSD. The situation when translational symmetry is enforced is, however, more subtle and the GSD of states with either $C$ even or odd can display anomalous ground state degeneracy that depends on the parity of $L_{x,y}$ as dictated by Eq.~(\ref{Eqn:AnomalousGSD}) and shown in Refs.~\onlinecite{Wen2003,Kou2008,Wen2009,Wen2010,Cho2012}. This will also be explicitly demonstrated with an exactly solvable model in Section~\ref{Sec:Model}. In fact the only states with even $C$ that have a consistent pattern of $\text{GSD}=4$ independent of $L_{x,y}$ are those with a completely trivial topological parity matrix $\zeta_{ij}=0$ which are adiabatically connected to the TC vacuum in the case of $C=0$. On the other hand, the only states with odd $C$ with a consistent pattern of $\text{GSD}=3$ independent of $L_{x,y}$ are those with a single non-trivial parity index $\zeta_{ij}=1$ and all others $\zeta_{ij}=0$, which are obtained from the those with $\zeta_{ij}=0$ by a single band inversion of the $\varepsilon$-fermions at a single special momenta $k_{ij}$, as discussed in the previous Section.

\subsection{Bulk-Edge Correspondence}

Another manifestation of the non-trivial weak topological invariants $\zeta_{ij}$ is the presence of dangling Majorana modes in open boundaries. Examples of this were presented in Refs.~\onlinecite{Wen2003,Kou2008,Wen2009,Wen2010,Cho2012}, but with our discussion it is possible to have a simple and systematic criterion for the appearance of dangling Majorana modes. Specifically, states where the $\varepsilon$-fermions form 2D stacks of Kitaev-wires will display an odd number of Majorana modes in exposed unit cells at some of the boundaries when translational symmetry along the boundary is preserved. In particular, in the basis for the topological indices described in the previous Section, $(C,\zeta_{K_x},\zeta_{K_y},\zeta_{\text{AI}})\in (\mathbb{Z},\mathbb{Z}_2,\mathbb{Z}_2,\mathbb{Z}_2)$, then we have that states with Kitaev-wire nature will have non-zero values of $(\zeta_{K_x},\zeta_{K_y})$, and will display an odd number of dangling Majorana modes in the corresponding boundaries. For example $(\zeta_{K_x},\zeta_{K_y})=(1,0)$ is a state with Kitaev-wires oriented along the $x$-direction and thus will have dangling Majorana modes along the exposed boundaries that are parallel to the $y$-direction. The Wen plaquette model,~\cite{Wen2003} which was the first example to be discovered of these anomalous states, is in fact topologically described by $(\zeta_{K_x},\zeta_{K_y})=(1,1)$, which means that it contains Kitaev-wires oriented along the diagonal and therefore displays dangling Majorana modes along both the $x$- and $y$-directions.

\subsection{Ideal Fixed Point Hamiltonians}\label{Sec:IdealHamiltonian}
In this Section we will construct ideal commuting projector Hamiltonians for all the phases with zero Chern number, namely those with $(C,\zeta_{K_x},\zeta_{K_y},\zeta_{\text{AI}}) =(0,\zeta_{K_x},\zeta_{K_y},\zeta_{\text{AI}})$. It is rigorously known that for phases with a U($1$) symmetry, so that the Chern number implies a non-zero Hall conductivity, it is impossible to construct local commuting projector Hamiltonians.~\cite{Kapustin2020} Presumably, this remains true in general whenever the spectral Chern number, $C$, is non-zero, regardless of whether the system has a U($1$) symmetry. Note however that this clearly does not imply that one cannot construct exactly solvable models of phases with non-zero $C$, as demonstrated by the Kitaev honeycomb model,~\cite{Kitaev2006} and as we will also illustrate in Section~\ref{Sec:Model}. The commuting projector Hamiltonians will, however, prove useful in illustrating the phenomenon of `weak breaking of translational symmetry',~\cite{Kitaev2006} associated with phases with non-trivial topological parity indices $(\zeta_{K_x},\zeta_{K_y},\zeta_{\text{AI}})$  that we will discuss in Section~\ref{Sec:WeakSymmetryBreaking}. Each of this phases can in turn be obtained as the ground state of a commuting projector Hamiltonian of the form:
\be\label{Eqn:IdealHamiltonian}
H = -\Delta_e \sum_v \Gamma^e_v -\Delta_\varepsilon \sum_p C_p.
\ee
Here $\Gamma^e_v$ is the parity of the $e$-particle, defined in Eq.~(\ref{Eqn:eDefinition}), and $C_p$ are $\mathbb{Z}_2$-valued operators ($C_p^2=1$) that act on a finite number of spins in the vicinity of plaquette $p$, and all operators in the Hamiltonian commute with each other:
\be
[C_p,C_{p'}]=0,~[C_p,\Gamma^e_v]=0.
\ee
The operator $C_p$ depends on the phase in question, labeled by parity indices $(\zeta_{K_x},\zeta_{K_y},\zeta_{\text{AI}})$, and we choose it so that under the fermion duality it maps onto a Majorana fermion bilinear of the form $C_p\leftrightarrow i\gamma_{1(p)}\gamma'_{2(p)}$ in the sector with no $e$-particles ($\Gamma^e_v=1$), and onto the corresponding fermion bilinear with twisted phases in the sectors with $e$ particles and non-trivial twists of boundary conditions, as described in Section~\ref{Sec:PeriodicLattice}. The Hamiltonians of Eq.~(\ref{Eqn:IdealHamiltonian}) will realize different phases depending on the sign of $\Delta_\varepsilon$, and these are listed in Table~\ref{Table:IdealOperators}. The detailed analysis to construct these operators in the case of the phases with diagonal stacking of Kitaev wires (KW$_{x+y}$ phases) is presented Appendix~\ref{Appendix:DiagonalStacking}. The pattern of Majorana pairing for each of these ideal Hamiltonians is illustrated in Figs.~\ref{Fig:HoneyComb} and \ref{Fig:WenPhase}, which makes clear the interpretation of a given phase as `atomic insulator' or a stack of Kitaev wires, and it is also straightforward to visualize which phases will have dangling Majorana modes in their boundaries.

\begin{table}
	\begin{tabular}{|c|c|c|c|}
		\hline
		phases &$C_p$&$\sign \Delta_\varepsilon$& Examples  \\\hline
		AI$_0$&$\Gamma^\varepsilon_p$&+& Toric Code~\cite{Kitaev2003} \\\hline
		AI$_1$&$\Gamma^\varepsilon_p$&-& \\\hline
		KW$_{x,0}$ & $U_{x,p}$& +& \\\hline
        KW$_{x,1}$ & $U_{x,p}$& -&  \\\hline
        KW$_{y,0}$ & $U_{y,p}$& +&  \\\hline
        KW$_{y,1}$ & $U_{y,p}$& -&  \\\hline
        KW$_{x+y,0}$ & $ \Gamma^\varepsilon_{NE(p)} U_{y,E(p)}U_{x,p} $& +& \\\hline
        KW$_{x+y,1}$ & $ \Gamma^\varepsilon_{NE(p)} U_{y,E(p)}U_{x,p} $& -& Wen model~\cite{Wen2003} \\\hline
	\end{tabular}
		\caption{\label{Table:IdealOperators}  $C_p$ for different phases for the ideal fixed-point Hamiltonian in Eq.~(\ref{Eqn:IdealHamiltonian}). $E(p)$ and $NE(p)$ are plaquettes to the east and north-east of plaquette $p$. Examples for the phases in which there is no entry under the "Examples" column are realized by the ideal Hamiltonian described in Sec.~\ref{Sec:Model}.}
\end{table}

\subsection{Weak Breaking of Translational Symmetry}\label{Sec:WeakSymmetryBreaking}
One of the most remarkable consequences of the non-trivial weak topological superconductivity of the $\varepsilon$-particles is the concomitant appearance of a phenomenon called `weak symmetry breaking' in Ref.~\onlinecite{Kitaev2006}. The idea is that, in certain topological phases, the action of a symmetry group can non-trivially exchange different anyon kinds (super-selection sectors).~\cite{Barkeshli2019} In the case of translational symmetry that we are studying, this manifests, for example, by a translation that maps an $e$-particle into an $m$-particle, as it occurs in the Wen plaquette model.~\cite{Wen2003} The underlying mechanism for why this phenomenon appears hand in hand with the GSD anomalies and the dangling Majorana modes, has not been described before, but as we will see, it is intimately tied to the formation of stacks of Kitaev-wire states by the $\varepsilon$-fermions. We will now discuss a systematic connection between patterns of weak symmetry breaking and the underlying topological indices,$(C,\zeta_{K_x},\zeta_{K_y},\zeta_{\text{AI}})\in \mathbb{Z}\times (\mathbb{Z}_2)^3$. To do so, we will exploit the ideal commuting projector fixed point Hamiltonians from Section~\ref{Sec:IdealHamiltonian}, but with the implicit idea that the results would carry over as universal properties of the phases they belong to. We recall from Section~\ref{Sec:OpenLattice} that we have enforced a local conservation law of an operator that measures the presence of the $e$-particles added on top of the TC vaccum, given in Eq.~(\ref{Eqn:eDefinition}). Let us consider a single $e$ particle placed in a vertex $v$ in an infinite lattice. The presence of this particle requires to twist the boundary conditions for the $\varepsilon$-fermions hopping across a line that extends from the vertex containing the $e$-particle towards infinity. Now, the pair-creation or transport operator of such $e$-particle between two nearby vertices $v_1$ and $v_2$, $T^e_{v_1v_2}$, will generally depend on the specific state the $\varepsilon$-fermions are in, but it must satisfy the following criteria:
\begin{enumerate}
	\item It should only create two $e$-particles on $v_1$ and $v_2$. Namely it should only anti-commute with the $e$-particle parities in the two vertices in question, $\Gamma^e_{v_1},~\Gamma^e_{v_2}$, and commute with the $e$-particle parities elsewhere.
	\item It should be local. Namely it only acts on physical spins within a certain finite radius of $v_1,v_2$ (for non-ideal Hamiltonians away from the commuting projector fixed point, it would have exponentially decaying overlap with distant spin operators).
	\item It should commute with the $C_p$ term of the ideal fixed point Hamiltonian in Eq.~(\ref{Eqn:IdealHamiltonian}). This is because when it transports an $e$-particle initially located at $v_1$ to the vertex $v_2$, both initial and final states should have the same energy in order for it to have the interpretation of an $e$-particle transport operator. (For non-ideal Hamiltonians away from the commuting projector fixed point, this should remain true in the limit of an infinite transitionally invariant lattice when the string of the single $e$-particle extends to infinity).
\end{enumerate}

Let us describe these transport operators first in the simplest phases, which are the atomic insulators AI$_0$ and AI$_1$. The ideal fixed point Hamiltonian for AI$_0$ is equivalent to the one of the usual Toric Code,~\cite{Kitaev2003} and for AI$_1$ it is that of the TC but with opposite sign for the plaquette opeator shown in Table~\ref{Table:IdealOperators}. Thus the $e$-particle pair-creation operators between two neighbouring vertices $v_1$ and $v_2$, $T^e_{v_1v_2}$ are simply given by:
\be
T^e_{v_1v_2}= Z_{v_1v_2},
\ee
where $Z_{v_1v_2}$ operates on the link connecting the two vertices. Notice that the operator that transports the $e$ particle over the smallest allowed closed loop (one plaquette), is simply $G^m_p$ and is algebraically \textit{dependent} on the operators appearing in the ideal fixed point Hamiltonian. This is a general property of any ideal fixed point Hamiltonian, since contractible closed loop transport operators must commute with the Hamiltonian, and therefore they cannot be algebraically independent of those appearing in the commuting projector Hamiltonian, since these provide a complete algebraic basis all local operators that commute with the Hamiltonian. Thus we see that the two vacua AI$_0$ and AI$_1$ are eigenstates of the closed loop transport operator of $e$-particles, but with opposite eigenvalues 1 and -1 respectively, reflecting the fact that the $e$-particles experience a background $\pi$-flux per plaquette in the AI$_1$ phase containing one $\varepsilon$-fermion per plaquette. Therefore, in the case of atomic insulator phases (AI$_i$), there is no weak symmetry breaking of translations, but instead there appears a projective representation of the translational symmetry group~\cite{Wen2002} of $e$-particles in the AI$_1$ phase, analogous to magnetic-translations with $\pi$-flux per unit cell.

\begin{figure}
	
	\includegraphics*[width=0.8\linewidth]{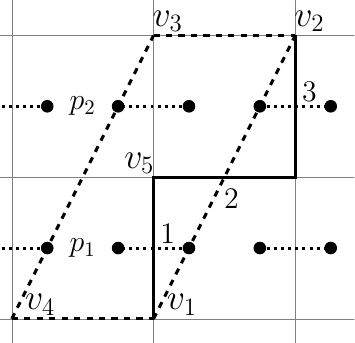}
	
	\caption{\label{Fig:WeakSymmetryA2} Emergent anyonic statistics of $e$-particles for KW$_{x,\zeta}$ phases. As shown in Eq.~(\ref{Eqn:AnyonTransport2}), a closed loop transport operator (dashed line) between odd-odd or even-even rows measures the $e$-parity of vertices contained. $e$ on odd and even rows are effective $e'$ (i.e. on vertices $v_{1-4}$) and $m'$ ($v_5$) of the Toric Code. Note hopping between adjacent rows will cut the Majorana bond (dotted line) odd times. The solid black line corresponds to hopping two rows given by Eq.~(\ref{Eqn:AnyonTransport}). }
\end{figure}

However, the situation changes considerably in the phases that have stacks of Kitaev wires of $\varepsilon$-fermions. To construct the transport operators in these cases, we begin by noticing that these phases generally break the $C_4$ rotational symmetry, and therefore, we expect the translation operators along the $x$- and $y$-directions to differ. We will illustrate this explicitly for the KW$_{x,\zeta}$ phases but similar considerations apply to the other phases that can be viewed as stacks of Kitaev wires. It is easy to verify that for the KW$_{x,\zeta}$ phase with Kitaev wires running along the $x$-direction, the $e$-particle pair creation operator remains the same as in the ordinary TC (AI$_0$ phase), for neighboring vertices along the $x$-direction. This is because the flux pair creation connecting nearest neighbor vertices does not intersect the bonds that pair Majorana modes in the given phase, as depicted in Fig.~\ref{Fig:WeakSymmetryA2}. In other words, moving the flux along the direction of the wires commutes with operators describing fermion hopping and pair-fluctuation, since it does not introduce branch-cuts along the bonds belonging to wires according to the principles described in Sections~\ref{Sec:Review} and \ref{Sec:PeriodicLattice}.

On the other hand, the operator that pair-creates $e$-particles in the TC vacuum for nearest neighbor vertices along the $y$-direction, which is orthogonal to the wires, does not commute with the $C_p$ term in Hamiltonian of Eq.~(\ref{Eqn:IdealHamiltonian}) for the KW$_{x,\zeta}$, and therefore violates the principle (3) of $e$-particle pair creation or transport operators. In fact, there is a fundamental obstruction to constructing an operator satisfying all of the three criteria that would transport a flux between nearest neighbor vertices that intersect one of the wires in the corresponding KW$_{x,\zeta}$ phase. To see this let us consider placing the system in a torus. Notice that if we hop a flux that initially resides say in vertex $v$ to the neighboring vertex $v+y$, then, in the final configuration, the $C_p$ operator of the bond that is intersected by such flux hopping would be mapped into a fermion bilinear with an extra minus, according to principles described in Sections.~\ref{Sec:Review} and \ref{Sec:PeriodicLattice} and illustrated as solid black line in Fig.~\ref{Fig:WeakSymmetryA2}. This implies that the intersected Kitaev wire would change boundary conditions under such flux hopping. However the ground state of a Kitaev wire with periodic boundary conditions has an even number of fermions, whereas the ground state with anti-periodic boundary conditions has an odd number of fermions. Therefore, the flux hopping would change the total $\varepsilon$-fermion parity of the system by $1$, which is not allowed in the torus. Therefore, from the above argument, we conclude that the only way to hop the flux across a single Kitaev wire would require the creation of one Bogoliubov fermion added on top of the vacuum with an energy cost of $\Delta_\varepsilon$, and thus would violate principle (3). In open boundary conditions it is possible to hop the flux across a single Kitaev wire, at the expense of adding a single $\varepsilon$-fermion (see Section~\ref{Sec:PeriodicLattice} for discussion on single fermion creation in open lattices), which would allow to satisfy criterion (3), but would violate the criterion (2), since the single fermion creation is necessarily non-local. We are thus led to the remarkable constraint that it is impossible to hop or pair create fluxes along neighboring vertices in the $y$-direction for KW$_{x,\zeta}$, while satisfying the three criteria above.

It is, however, possible to pair-create (or hop) $e$-particles that are second nearest neighbor vertices along the $y$-direction for KW$_{x,\zeta}$, while satisfying all the 3 criteria as illustrated in Fig.~\ref{Fig:WeakSymmetryA2}. The operators accomplishing this for the KW$_{x,\zeta}$ phase are given by:
\be\label{Eqn:AnyonTransport}
T^e_{v_1v_2}= i Z_2 Z_3 X_2= i (Z_1Z_2Z_3) \times (X_2Z_1),
\ee
shown visually as solid black line in Fig.~\ref{Fig:WeakSymmetryA2}. In the last equality of Eq.~(\ref{Eqn:AnyonTransport}), we have written the transport operator as a product of the `bare' $e$-transport operator in the TC (product of $Z$s) and a vertical Majorana pair creation operator [$U_{p,y}$ from Eq.~(\ref{Eqn:PairCreation})]. The reason this is possible is that when hopping an $e$-particle across two Kitaev wires, one twists the boundary condition of both neighboring wires, and, therefore, if one would use the bare hopping operators of $e$-particles from the TC vacuum, one would have two Bogoliubov fermions added to each of these wires in the two bonds that are intersected by such hopping. These Bogoliubov fermions, however, can be destroyed locally by a Majorana bilinear operator that connects the adjacent wires, restoring both wires back to their ground states with the twisted boundary conditions that are induced by the $\varepsilon$-particle hopping.

From the operators that produce the smallest allowed hoppings of $e$ particles in the KW$_{x,\zeta}$ (KW$_{y,\zeta}$) phases, given in Eq.~(\ref{Eqn:AnyonTransport}), it is possible to then construct the operator that moves the $e$-particles around the smallest allowed closed loop (depicted in Fig.~\ref{Fig:WeakSymmetryA2}). This operator can be interpreted as creating two pair of particles in neighbouring vertices and then annihilating one pair after completing the smallest allowed closed loop transport of $e$-particles. Therefore this operator must commute with the ideal fixed point Hamiltonian from Eq.~(\ref{Eqn:IdealHamiltonian}) of the corresponding phase. For the KW$_{x,\zeta}$ phases the closed-loop transport operator is given explicitly by:
\be\label{Eqn:AnyonTransport2}
\prod_{v} T^e_{v_iv_j}=\Gamma^e_{v_5} C_{p_1}C_{p_2},~C_p = U_{x,p},
\ee
where the path is shown as the dashed line in Fig.~\ref{Fig:WeakSymmetryA2}. Notice the appearance of $\Gamma^e_{v_5}$ in Eq.~(\ref{Eqn:AnyonTransport2}). This implies that the closed transport of $e$-particles in the smallest allowed loop for the phase KW$_{x,\zeta}$ equals the identity in the ground state, but there is a non-trivial semionic statistic among $e$-particles that belong to the vertices that are separated by a single Kitaev wire and that cannot be connected by any local $e$-particle transport operator. Therefore we are led to the remarkable conclusion that the $e$ particles in these two kinds of vertices, are distinct anyons with mutual semionic statistics that belong to two different super-selection sectors. 

All of the above conclusions apply as well to the phases KW$_{y,\zeta}$ and KW$_{x+y,\zeta}$, which can be viewed as having stacking of Kitaev wires along vertical and diagonal directions. In the case of KW$_{x+y,\zeta}$ phases, the vertices that can be connected belong to the two sub-lattices of the square lattice. Details of the transport operators in this case are presented in Appendix~\ref{Appendix:DiagonalStacking}.

Let us then summarize the picture that emerges from the above considerations for the phases that can be viewed as stacks of Kitaev wires of $\varepsilon$-fermions. The $e$-particles in these phases are separated into two super-selection sectors. $e$-particles in vertices separated by crossing an even (odd) number of Kitaev wires belong to same (different) super-selection sector. The above is the phenomenon of weak symmetry breaking, as introduced in Ref.~\onlinecite{Kitaev2006}. These two kinds of $e$-particles of different super-selection sectors have the same bulk topological properties of the $e$ and $m$ particles of an ordinary TC. In other words, even when we force the original $\varepsilon$-fermions of the toric code to not appear at low energies [say by taking $\Delta_\varepsilon$ to be large and positive in Eq.~(\ref{Eqn:IdealHamiltonian})], there is an emergent anyon statistics of the fluxes in such background of gapped fermionic matter, forced upon them by the topology of the underlying Kitaev wires.

\section{Model}\label{Sec:Model}

The results in Section~\ref{Sec:Review} allow us to construct a large class of exactly solvable spin models of $\mathbb{Z}_2$ topologically ordered states, one for each free fermion Hamiltonian. In this Section we will illustrate this in a specific model [Eq.~(\ref{Eqn:Decoupled}) below], which will realize $14$ out of the $16$ classes of states with non-trivial parity indices given in Section~\ref{Sec:TopologicalSuperconductor}. Moreover, the model contains $6$ out of the $8$ topological phases of the ideal fixed-point Hamiltonian; see Section~\ref{Sec:IdealHamiltonian}. As we will see, some of these phases will feature anomalous GSD that depends on the size of the torus, and some will feature dangling Majorana modes in open boundaries, in line with the considerations of Section~\ref{Sec:TopologicalSuperconductor}, and we will be able to provide exact solutions for both their bulk and boundary spectrum.

\begin{figure*}
	\includegraphics*[width=0.3\linewidth]{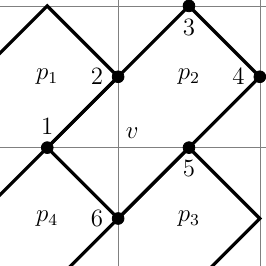}\llap{
		\parbox[b]{11cm}{\large (a)\\\rule{0ex}{4.7cm}
	}}~~
	\includegraphics*[width=0.6\linewidth]{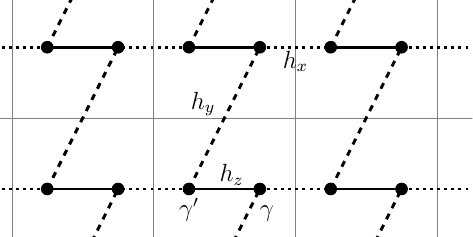}\llap{
		\parbox[b]{20.4cm}{\large (b)\\\rule{0ex}{4.7cm}
	}}
	\caption{\label{Fig:HoneyComb} (a) Visual representation of Eq.~(\ref{Eqn:HoneyCombOperators}) as a distorted `honeycomb' lattice. (b) Equivalence of Eq.~(\ref{Eqn:Decoupled1}) to a Kitaev honeycomb model. $\gamma'$ and $\gamma$ are defined on left and right sides of each plaquette. Couplings between $\gamma', \gamma$ in Eq.~(\ref{Eqn:Decoupled1}) are illustrated by: solid lines for fermion parity $\Gamma^\varepsilon$; dotted lines for horizontal hopping $U_x$; dashed lines for vertical hopping $U_y$. }
\end{figure*}

We choose the Hamiltonian to be:
\begin{equation}
\begin{split}\label{Eqn:Decoupled}
&H = H_0 + V,\\
H_0=&-\Delta_e\sum_v  \Gamma^e_v  - \sum_p\bigg( h_x U_{x,p} + h_y U_{y,p} + h_z \Gamma^\varepsilon_p  \bigg),\\
V =&  \frac{i\delta}{2} \sum_p \bigg[U_{y,p} \bigg(\Gamma^\varepsilon_p + \Gamma^\varepsilon_{N(p)} \bigg)\bigg],~\delta,\Delta_e>0.
\end{split}
\end{equation}
$N(p)$ is the plaquette to the north of $p$. This Hamiltonian conserves the local parity of $e$-particles at each vertex, measured by $\Gamma^e_v$. We will be interested in excitations belonging to the sector without $e$-particles, which energetically can be enforced to be the ground state sector by assuming that $\Delta_e\gg |h_{x,y,z}|, |\delta|$. Therefore, this Hamiltonian can be exactly mapped into a dual local fermionic Hamiltonian even in geometries with open boundaries such as the cyclinder or the open lattice described in Section~\ref{Sec:Review}, via Eqs.~(\ref{Eqn:OperatorDefinitions}) and (\ref{Eqn:BoundaryHopping}).

As we will see, the Hamiltonian from Eq.~(\ref{Eqn:Decoupled}) maps exactly into a free fermion bilinear Hamiltonian for any values of its parameters and it is therefore generally exactly solvable. For $h_x=h_y=\delta=0$ and $h_z>0$, this model is equivalent to the Toric code.~\cite{Kitaev2003} Additionally, for $\delta=0$, this model is equivalent to the Kitaev honeycomb model in the sector with no fluxes, $\Gamma^e_v=1$, for all $v$.~\cite{Kitaev2006} More precisely, the following operators are unitarily equivalent to two-spin operators in Kitaev's honeycomb model in all sectors regardless of $\Gamma^e_v$, which we show visually in Fig.~\ref{Fig:HoneyComb}: 
\be\label{Eqn:HoneyCombOperators}
U_{x,p_1} = X_2 Z_1,~ U_{y,p_2}= X_3 Z_2,~U_{x,p_2}\Gamma^\varepsilon_{p_2} U_{y,p_2}= Y_3 Y_4.
\ee
It follows that, after a unitary transformation on points $2, 5$, and by viewing the lattice as a honeycomb, as depicted in Fig.~\ref{Fig:HoneyComb}, we recover $x$-, $y$- and $z$-links of the Kitaev honeycomb model. $\Gamma^e_v$ is then mapped to the plaquette operator $W_{p_2}$ to its north-east. Unless otherwise noted, throughout this work we will view the geometry of this model as that of a square lattice rather than a honeycomb.

In Section~\ref{Sec:PhaseDiagram} we consider the Hamiltonian on an infinite lattice and study the general phase diagram in the parameter space of $(h_x/|h_z|,h_y/|h_z|)$ and $\delta>0$. Its properties in a finite torus and in open lattices will be discussed in Sections~\ref{Sec:Torus} and \ref{Sec:OpenBoundary}, demonstrating its anomalous GSD and its gapless boundary Majorana modes. 


\subsection{Infinite Lattice}\label{Sec:PhaseDiagram}

\begin{figure*}
	
	\includegraphics*[width=0.4\linewidth]{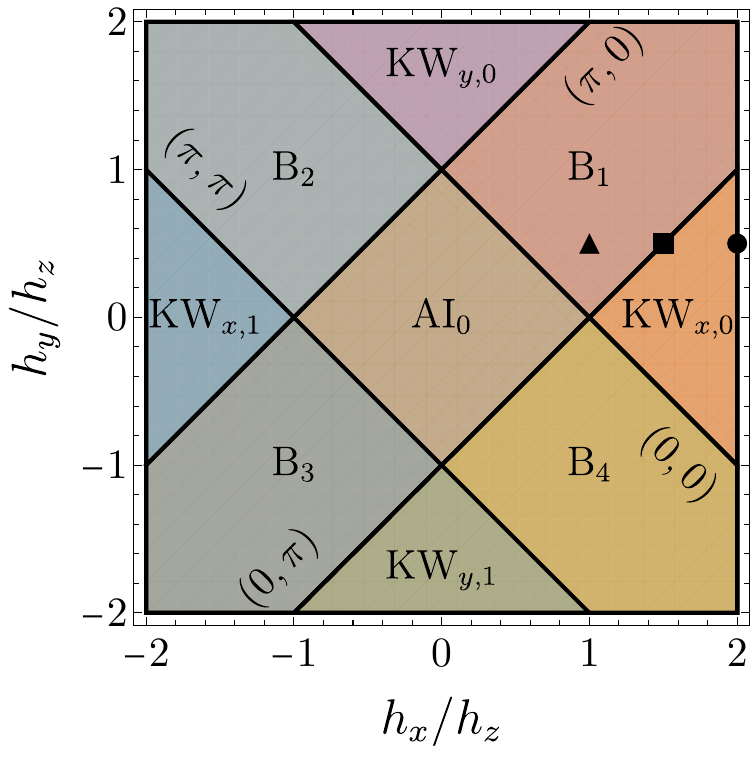}\llap{
		\parbox[b]{13.5cm}{\large (a)\\\rule{0ex}{6.5cm}
	}}~~
	\includegraphics*[width=0.4\linewidth]{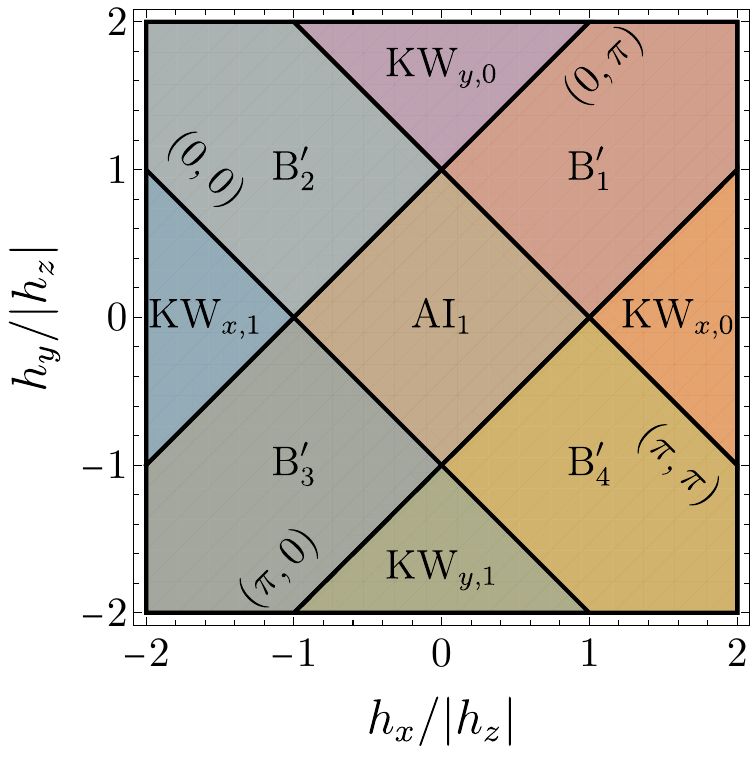}\llap{
		\parbox[b]{13.5cm}{\large (b)\\\rule{0ex}{6.5cm}
	}}~~
	
	\caption{\label{Fig:PhaseDiagram} Phase diagrams for (a) $h_z>0$ and (b) $h_z<0$ respectively. The 2-tuple $(k_x,k_y)$ near each critical line is the momenta of gap closing at that critical line. }
\end{figure*}

On an infinite square lattice, the Hamiltonian from Eq.~(\ref{Eqn:Decoupled}) can be mapped directly into a sum of fermion bilinears. Substituting Eqs.~(\ref{Eqn:OperatorDefinitions}) and (\ref{Eqn:MajoranaFermions}) into Eq.~(\ref{Eqn:Decoupled}) leads to:
\begin{multline}\label{Eqn:Decoupled1}
H= - \sum_{i,j} \bigg(h_x a_{i,j}^\dagger a_{i,j+1} +h_y a_{i,j}^\dagger a_{i+1,j} -  h_z a_{i,j}^\dagger a_{i,j}+ \\ +h_x a_{i,j} a_{i,j+1} + h_y a_{i,j} a_{i+1,j} \bigg)-i\delta \sum_{i,j} a_{i,j}a_{i+1,j}+\text{h.c.}.
\end{multline}
Here $i, j$ are row and column indices of a given plaquette. Notice that the pairing terms in Eq.~(\ref{Eqn:Decoupled1}) respect translational symmetry, and, therefore, Eq.~(\ref{Eqn:Decoupled1}) has the form of a mean-field BCS fermion bilinear Hamiltonian with zero center-of-mass momentum for Cooper pairs. We split each of the complex fermions operators at a given site into two Majorana operators using Eq.~(\ref{Eqn:MajoranaFermions}):
\be
a= \frac{1}{2}(\gamma+i\gamma'),~a^\dagger= \frac{1}{2}(\gamma-i\gamma').
\ee
The Hamiltonian in Eq.~(\ref{Eqn:Decoupled1}) can be visualized by regarding each $\gamma, \gamma^\prime$ as Majorana fermion modes residing on plaquettes of the square lattice, and viewing $h_x, h_y, h_z$ as bond dependent Majorana pairing terms in the lattice, as depicted in Fig.~\ref{Fig:HoneyComb}. As mentioned before, this model is equivalent to Kitaev honeycomb model,~\cite{Kitaev2006} although the fermionic duality described in Section~\ref{Sec:Review} allows one to solve the Hamiltonian without explicitly enlarging the local Hilbert space, and this is why there are only $2$ Majorana modes per plaquette, which are sufficient to exhaust all the local degrees of freedom in the sector with no flux. Also, we have added an explicit energy cost, $\Delta_e$, to gap the $\mathbb{Z}_2$ fluxes ($e$-particles) to make sure they are not part of the ground state sector of interest. The phase diagram is equivalent to the one in Ref.~\onlinecite{Kitaev2006} for the case $\delta=0$, and it is shown in Fig.~\ref{Fig:PhaseDiagram}. The gapless phases are $B_{1-4},~B'_{1-4}$ while the other phases are gapped. In particular, phases $B_1, \text{AI}_0, \text{KW}_{x,0}, \text{KW}_{y,0}$ are $B, A_y, A_x, A_z$ in the Kitaev Model. With a finite $\delta$, $V$ acts as second nearest neighbour hopping $i\delta (\gamma \gamma -\gamma'\gamma')$ along the vertical direction only. It is similar, but not identical, to the perturbation induced by the magnetic field in Ref.~\onlinecite{Kitaev2006}, which couples all the second nearest neighbors, but it produces essentially the same effect in that $V$ gaps all gapless phases without shifting the phase boundaries. For the remainder, in order to ensure that all the phases are gapped so that they can be classified within the scheme described in the previous Section, we will fix $\delta>0$ unless otherwise stated. This also allows to associate a Chern number to each phase; see Appendix~\ref{Appendix:ChernNumber}.

Let us compute the BdG spectrum of this Hamiltonian. Going over to momentum space using the convention of the square lattice (which differs from the honeycomb) $a_{i,j}=\sum_{\bm{k}} a_{\bm{k}} \exp(i\bm{k}.\bm{r}_{ij})$, Eq.~(\ref{Eqn:Decoupled1}) becomes Eq.~(\ref{Eqn:BCSHamiltonian}) with $\varepsilon(\bm{k}) = -2(h_x\cos k_x +h_y \cos k_y - h_z)$ and $\Delta(\bm{k})=2\delta \sin k_y- 2i(h_x \sin k_x - h_y \sin k_y)$. The lattice constant is set to unity. The dispersion of Bogoliubov fermions is:
\begin{multline}\label{Eqn:Spectrum}
E(\bm{k})=\pm \bigg[4\bigg(h_x\cos k_x +h_y \cos k_y - h_z\bigg)^2+|\Delta(\bm{k})|^2\bigg]^{\frac{1}{2}}.
\end{multline}
From the above dispersion, one can show that all of the phases are in fact separated by a critical line at which the Bogoliubov spectrum becomes gapless at some special momentum in the BZ of the square lattice. Therefore, one can obtain the phase diagram by solving for $E(\bm{k}) = 0$ and the phases are shown in Fig.~\ref{Fig:PhaseDiagram}. The critical lines separating different phases are labeled by the `high-symmetry' momentum points $\bm{k}_0$ where the dispersion is gapless. As outlined in Section~\ref{Sec:TopologicalSuperconductor}, these phases are classified by the four parity labels at these momenta, and the Chern number subject to constraint (\ref{Eqn:ChernNumberConstraint}). We note that the model includes $6$ out of the $8$ phases in Section~\ref{Sec:IdealHamiltonian} with trivial Chern number ($C=0$) which can be viewed as lower dimensional stacks of $\varepsilon$-fermion wires, since the lattice Hamiltonian in Eq.~(\ref{Eqn:Decoupled}) approaches the corresponding ideal fixed-point Hamiltonians in certain limits of the parameter space. A model for the two remaining phases that are not realized by this model, namely KW$_{x+y,\zeta}$ phases is constructed in Appendix~\ref{Appendix:DiagonalStacking}.

When we also include the phases with finite Chern number, the current model realizes a total of $14$ topologically distinct phases, when they are viewed as topological phases enriched by translational symmetry. Some of these phases can be distinguished by the topological characteristics of its bulk excitations without any regard to symmetry, in the same spirit of the Kitaev 16-fold classification, namely, they can be distinguished by the spectral Chern number $C$ of the BdG spectrum.~\cite{Kitaev2006} In our model, the Chern numbers of these $14$ phases are:
\be\label{Eqn:ChernNumber}
\begin{split}
&C=0:\text{AI}_0,~\text{AI}_1,~\text{KW}_{x,0},~\text{KW}_{x,1},~\text{KW}_{y,0},~\text{KW}_{y,1};\\
&C=1:\text{B}_1, \text{B}_2,~\text{B}_3',~\text{B}_4';\\
&C=-1:\text{B}_3,~\text{B}_4,~\text{B}_1',~\text{B}_2'.
\end{split}
\ee
Details of calculations are given in Appendix~\ref{Appendix:ChernNumber}. From above one might naively think that, since phases such as AI$_0$ (Toric Code) and KW$_{x,0}$ have the same Chern number $C=0$, the gap closing along the line $h_x=h_z, h_y=0$ might be accidental and could be removed by adding a perturbation, so that the ground states in region AI$_0$ could be deformed adiabatically into those in region KW$_{x,0}$. In fact, some of these phases can in a sense be recast exactly as Toric code models in certain limits in infinite lattices or in periodic lattices with an even number of Kitaev wires, as shown in Appendix~\ref{Appendix:ToricCodeGS}. However, these phases can be distinguished by the topological parity indices described in Section~\ref{Sec:TopologicalSuperconductor}, and therefore, provided the underlying translational symmetry is preserved, they are necessarily separated by an intermediate gapless critical phase.

Let us now determine the matrix of fermion parity at special momenta, $\zeta_{ij}$, discussed in Section~\ref{Sec:TopologicalSuperconductor}, for these phases. Following Eq.~(\ref{Eqn:TopologicalIndex}), the parity can be simply determined by sign of the diagonal part of the BdG Hamiltonian for a single orbital model, which for the Hamiltonian from Eq.~(\ref{Eqn:Decoupled1}), reads as:
\be
\varepsilon(\bm{k}) = -2(h_x\cos k_x +h_y \cos k_y - h_z).
\ee
Direct calculations show that the topological parity matrices, $\zeta_{ij}$, in the convention of Eq.~(\ref{Eqn:MomentaMatrix}), for the phases with $C=0$ listed in Eq.~(\ref{Eqn:ChernNumber}) are given by the matrices listed in Eqs.~(\ref{Eqn:Phases1})-(\ref{Eqn:Phases4}), and this is why we have labeled them accordingly. In fact, these phases realize the fixed point ground states of commuting projector Hamiltonians discussed in Section~\ref{Sec:IdealHamiltonian} in the appropriate limits. For the KW$_{x,\zeta}$ phases the fixed point is realized by setting $h_y=h_z=\delta=0$ and $\zeta=0$ (1) corresponds to $h_x>0$ ($h_x<0$). Similarly AI$_\zeta$, and KW$_{y,\zeta}$ fixed points are realized by setting $h_x=h_y=\delta=0$ and $h_x=h_z=\delta=0$ respectively, and $\zeta$ is determined by the sign of the remaining non-zero $h_x$ or $h_z$ term.

For the phases with $C=\pm1$ listed in Eq.~(\ref{Eqn:ChernNumber}), we can similarly compute the parity indices and obtain:
\begin{eqnarray}
&\text{B}_1:\begin{pmatrix} 0 &0\\ 1& 0\end{pmatrix},~\text{B}_2:\begin{pmatrix} 1 &0 \\ 0& 0\end{pmatrix},~\text{B}_3:\begin{pmatrix} 0 &1 \\ 0& 0\end{pmatrix},
~\text{B}_4:\begin{pmatrix} 0 &0 \\ 0& 1\end{pmatrix},\nonumber \\
&\text{B}_1':\begin{pmatrix} 1 &0 \\ 1& 1\end{pmatrix},~\text{B}_2':\begin{pmatrix} 1 &1 \\ 1& 0\end{pmatrix},
~\text{B}_3':\begin{pmatrix} 1 &1 \\ 0& 1\end{pmatrix},
~\text{B}_4':\begin{pmatrix} 0 &1 \\ 1& 1\end{pmatrix}. \nonumber
\end{eqnarray}
These can be viewed as phases that are topologically equivalent to `layer addition' of the elementary phase with non-trivial Chern number, $\chi_C$ from Eq.~(\ref{Eqn:ChernNumber2}), and the phases that can be viewed as stacks of 1D wires. The two cases which are not realized in our model are the two KW$_{x+y,\zeta}$ phases which are in the same class of the Wen plaquette model,~\cite{Wen2003} and correspond to weak topological superconducting phases with diagonal stacking of Majorana wires. We describe exactly solvable models for these in Appendix~\ref{Appendix:DiagonalStacking}.

\subsection{Torus}\label{Sec:Torus}


Let us consider placing the Hamiltonian in Eq.~(\ref{Eqn:Decoupled}) on a square Torus with $L_{x,y}$ along the $x, y$-directions. Remarkably, the GSD may depend on $L_x, L_y$ being even or odd, as first pointed out in the example identified by Wen in Ref.~\onlinecite{Wen2003}. For example, for KW$_{x,\zeta}$ phases, the GSD degeneracy is 2 for $L_y$ odd but 4 for $L_y$ even. And for KW$_{y,\zeta}$, the GSD degeneracy is 2 for $L_x$ odd but 4 for $L_x$ even. Such GSD can be computed from Eq.~(\ref{Eqn:AnomalousGSD}). In Appendix~\ref{Appendix:Duality}, this computation of GSD is performed by mapping the system to a dual bosonic Hilbert space. Another method for performing this computation by directly counting constraints in the underlying spin degrees of freedom is also presented in Appendix~\ref{Appendix:GSDSpinLattice}.

We will now approach this phenomena by using the fermionic representation described in previous Sections and discuss the subtle interplay of the lattice size and the GSD in the torus geometry for the phases characterized by the aforementioned $\mathbb{Z}_2$ topological parity matrices; see Eq.~(\ref{Eqn:AnomalousGSD}). As is discussed in Section~\ref{Sec:Review}, in the torus geometry only states with an even number of fermions are physical, and therefore the physical GSD of a given phase depends on lattice size and fermion boundary conditions. As one moves from phase AI$_0$ (the Toric Code vacuum), which has no fermions in the ground state, the ground state $\varepsilon$-fermion parity changes upon crossing a critical line if the $\bm{k}_0$ at which the BdG gap closes is actually allowed for a given system size and boundary conditions. This GSD for any given phase can be computed explicitly using the formula from Eq.~(\ref{Eqn:AnomalousGSD}).

Consider, for example, phases B$_1$ and KW$_{x,0}$. As we start from AI$_0$(Toric Code), the phase transition onto B$_1$ occurs by closing the BdG gap at $\bm{k}=(0,0)$, and therefore the ground state in phase B$_1$ is forbidden for periodic boundary conditions along $x$-, $y$-directions, since $\bm{k}_0=(0,0)$ is allowed for any $L_{x,y}$. Since phase B$_1$ has Chern number $C=1$, it can be viewed to be topologically equivalent to the weak pairing phase of a 2D $p+ip$ spinless superfluid. The fact that these states have an odd number of fermions in the torus for periodic boundary conditions was first identified by Read and Green in their seminal work in Ref.~\onlinecite{Read2000}. Phase KW$_{x,0}$ however is a paired state which has Chern number $C=0$, but it still displays a nontrivial pattern of GSD depending on the system size. To see this, notice that in passing from B$_1$ to KW$_{x,0}$, the gap closes at $\bm{k}_0=(0,\pi)$. However, $\bm{k}_0=(0,\pi)$ is only part of the momentum lattice for periodic boundary conditions for $L_y$ even. Therefore, for periodic boundary and lattices with $L_y$ odd, the corresponding ground state of phase KW$_{x,0}$ has still the same parity as phase B$_1$, namely, an odd number of fermions, in spite of having a trivial Chern number. Similarly, for periodic and anti-periodic boundary conditions along $x$-, $y$-directions, $\bm{k}_0=(0,0)$ is always forbidden and $\bm{k}_0=(0,\pi)$ is allowed for $L_y$ odd. To have even total fermion parity for phase KW$_{x,0}$, $L_y$ must be even in both cases. For anti-periodic boundary condition along $x$-direction, both $\bm{k}_0=(0,0), (0,\pi)$ are not allowed, since $k_{0x}=0$ is not admitted, and phase KW$_{x,0}$ always has an allowed parity even ground state. This agrees with the alternative counting procedures presented in the underlying spin Hilbert space presented in Appendix~\ref{Appendix:GSDSpinLattice} that only anti-periodic boundary conditions for fermions along the $y$-direction is allowed for $L_y$ odd.

\begin{figure*}
	
	\includegraphics*[width=0.33\linewidth]{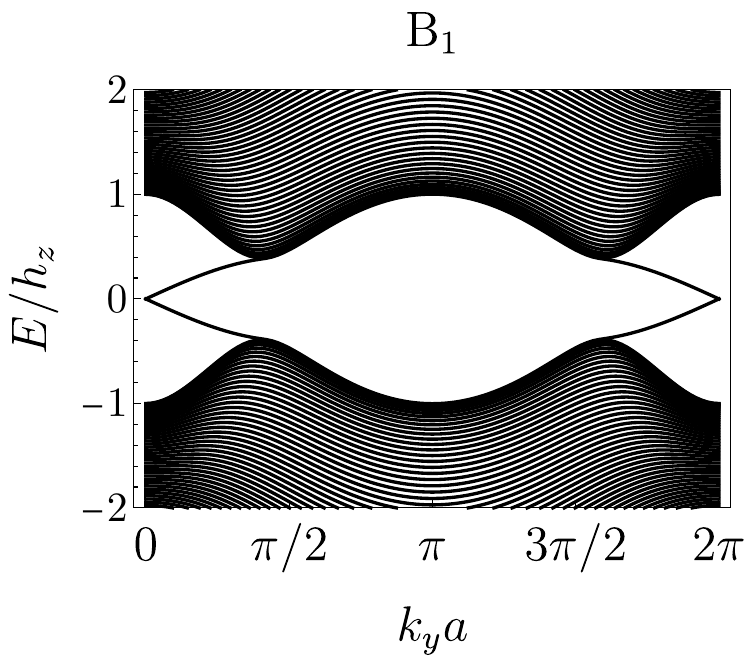}\llap{
		\parbox[b]{11cm}{\large (a)\\\rule{0ex}{4.7cm}
	}}%
	\includegraphics*[width=0.33\linewidth]{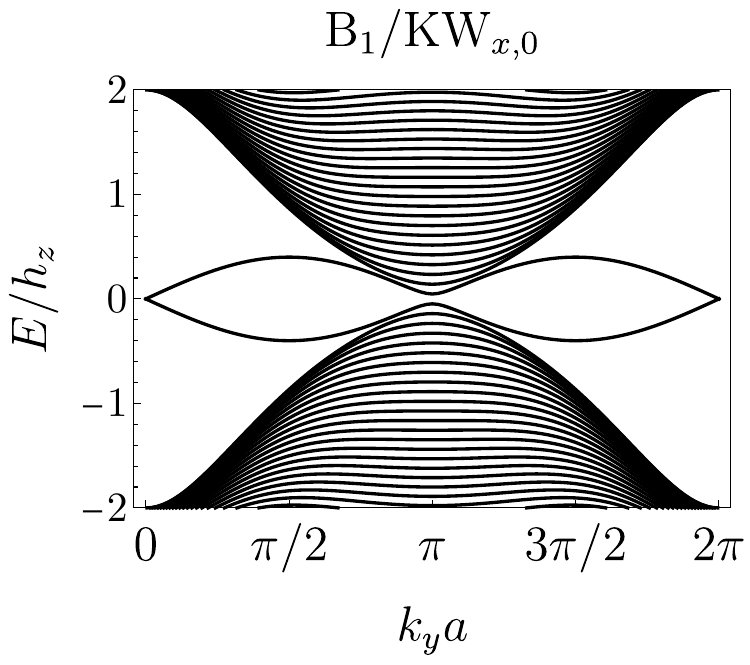}\llap{
		\parbox[b]{11cm}{\large (b)\\\rule{0ex}{4.7cm}
	}}%
	\includegraphics*[width=0.33\linewidth]{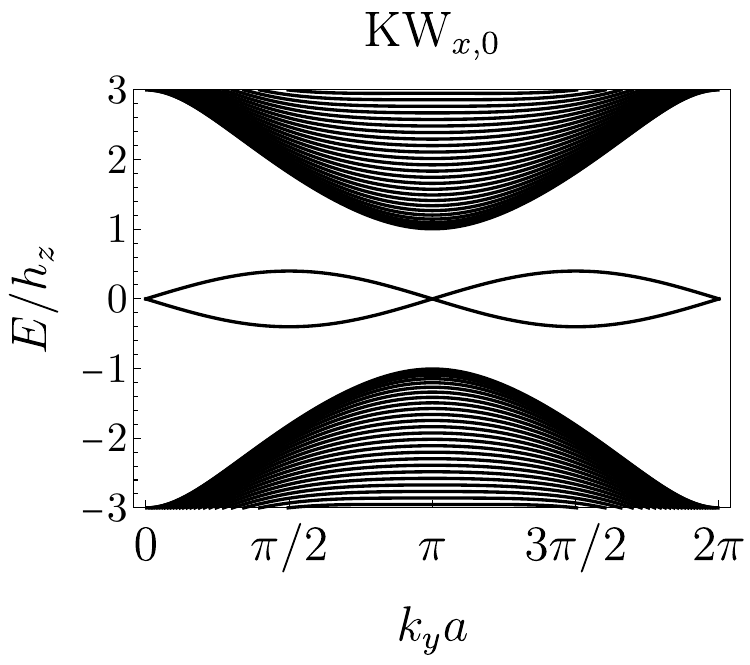}\llap{
		\parbox[b]{11cm}{\large (c)\\\rule{0ex}{4.7cm}
	}}%
	
	\caption{\label{Fig:EdgeModes} Edge Modes in a cylinder with the $y$-direction periodic and $L_x=100$. $h_y=0.5 h_z, \delta=0.2 h_z$ and $h_z>0$. (a)-(c) $h_x = 1,~1.5,~2 h_z$. (a) $h_x= h_z$ and (c) $h_x=2h_z$ belong to phases B$_1$ and KW$_{x,0}$. (b) $h_x=1.5 h_z$ is at the critical line between phases B$_1$ and KW$_{x,0}$. In Fig.~\ref{Fig:PhaseDiagram}, their locations in the phase diagram are marked with triangle, square and circle respectively. The gapless mode acquires the $k_y$ momentum of each critical line as one crosses from phase AI$_0$ into other regions in the phase diagram. The small splitting of zero modes at $k_y=\pi$ is a finite size effect.}
\end{figure*}

\subsection{Open and Cylindrical Lattices}\label{Sec:OpenBoundary}

We now consider open and cylindrical lattices to illustrate that topologically non-trivial phases in Section~\ref{Sec:Torus} feature gapless edge modes. The existence of chiral edge modes is not surprising for the B phases, since they have a non-zero $C=\pm 1$ spectral chern number, and therefore have a robust protected chiral Majorana edge mode. As we will see, some of the phases with $C=0$ in Eq.~(\ref{Eqn:ChernNumber}) have gapless edge modes that are not fully chiral, but still protected in the sense of the finer topological classification of their bulk based on the $\mathbb{Z}_2$ parity indices described in Section~\ref{Sec:Torus}.

In open and cylindrical lattices, there are no global constraints relating any of the elementary operators that make up the Hamiltonian in Eq.~(\ref{Eqn:Decoupled}). In the special limit in which the Hamiltonian reduces to the ideal commuting projector fixed point and the bulk has strictly flat bands with no dispersion, the existence of edge modes can be elucidated via a counting of degrees of freedom in the underlying spin Hilbert space for all of the phases that have zero Chern number in Eq.~(\ref{Eqn:ChernNumber}). For example, in the special case of the phase KW$_{x,0}$ whose fixed point Hamiltonian is realized for $h_z=h_y=\delta=0$ (which can be viewed as taking the limit $h_x\gg h_y,h_z,\delta$ in the phase diagram of Fig.~\ref{Fig:PhaseDiagram}) and open boundary conditions along $x$-direction, the two terms appearing in the Hamiltonian of Eq.~(\ref{Eqn:Decoupled}) $\Gamma^e$ and $U_x$ are commutative. There are $L_xL_y$ independent $\Gamma^e$ operators and $(L_x-1)L_y$ independent $U_x$ operators, since there is no hopping at the last column, as shown in Fig.~\ref{Fig:TCBoundary}. Since $\Gamma^e$ and $U_x$ are $\mathbb{Z}_2$ valued operators with eigenvalues $\pm1$, the number of subspaces of the Hilbert that can be labeled by distinct eigenvalues of these operators is then $2^{2 L_x L_y-L_y}$. However the total dimensionality of the underlying spin Hilbert space is $2^{2L_x L_y}$ and therefore each of these subspaces must be $2^{L_y}$ degenerate. In the fermionic representation it is easy to see that this degeneracy stems from isolated dangling Majorana modes along the vertical edges with $\sqrt{2}$ degrees of freedom per exposed plaquette on each of the open boundaries. This fact can be seen by going over to the dual fermionic Hilbert space. In the fermion representation, $U_{x,y}$ pairs Majorana modes $\gamma, \gamma'$ across plaquettes (see Fig.~\ref{Fig:HoneyComb}). The zero energy states are associated with the Majorana modes that remain unpaired in the exposed plaquettes at the open boundaries. They have zero energy since they commute with the fermionic Hamiltonian. For example, for $h_y=0$, these states are located along the vertical edges and are $\gamma_{1,n}$ on the first column and $\gamma_{L_x,n}$ on the last along a given row $n$. These are the zero edge Majorana dangling modes mentioned above.

Now, one of the great powers of the fermionic representation that we have developed for open and cylindrical lattices in Sections~\ref{Sec:OpenLattice} and \ref{Sec:CylindricalLattice}, is that it allows to obtain the exact eigenstates in these geometries even away from the ideal fixed point limit that leads to flat bands. In such cases the dangling Majorana modes that we just described are allowed to couple to form a non-trivially dispersing edge mode. For convenience we will present results only for the cylinder geometry, which can be more easily visualized since one direction remains fully translationally invariant, and thus quasi-1D dispersions can be plotted, although calculations in an open finite lattice are easily doable as well following the construction from Section~\ref{Sec:OpenLattice}. Assuming periodic boundary conditions along the $y$-direction, we partially Fourier transform the Majorana fermions along the $y$-direction and calculate the exact band-structure of Eq.~(\ref{Eqn:Decoupled}) for a large system size. The results are shown in Fig.~\ref{Fig:EdgeModes} for $L_x=100,~h_z=1, h_y=1/2$ and $h_x=1,~3/2,~2$. We see that, starting from phase AI$_0$, as each critical line is crossed, the spectrum acquires two Majorana modes with the corresponding $y$-component of momentum $\bm{k}_0$.

We now comment on the robustness of KW phases on an open lattice in relation to results obtained in the torus. As is shown in Section~\ref{Sec:PhaseDiagram}, for $L_y (L_x)$ odd, bulk orders of KW$_x$ phases (KW$_y$) on a Torus are stable with respect to perturbations and cannot be deformed adiabatically into phase AI$_0$ due to their distinct GSDs. On an open lattice, the stability of these phases manifests in the robustness of gapless Majorana modes with respect to local perturbations, and the same conclusion as in the Torus case holds. This can be seen for example in the case of the KW$_x$ phase which can be viewed as a stack of $L_y$ Kitaev wires of $\varepsilon$-fermions oriented in the $x$-direction. In the case of $L_y$ odd, it is impossible to gap all the Majorana modes, since there is an odd number of them in each edge, and there will always be an exact zero mode localized in each boundary of the cylinder.

\section{Summary and Outlook}\label{Sec:Conclusions}

In this work we have provided a unifying description of the interplay of topological order and translational symmetry in fractionalized states of matter with emergent $\mathbb{Z}_2$ gauge fields. We do this by exploiting the Toric Code as a convenient vacuum to construct states. Specifically, by enforcing a local symmetry which freezes the motion of isolated $e$- and $m$-particles, but allows the fluctuations of their fermionic bound state, the $\varepsilon$-particle, the underlying spin Hilbert separates into subspaces of $\varepsilon$-fermions coupled to non-dynamical background gauge fields. As recently emphasized in Ref.~\onlinecite{Chen2018}, this construction can be viewed as a form of two-dimensional Jordan-Wigner transformation or a type of charge-flux attachment that preserves spatial locality. We have elucidated this construction in geometries with fully open boundaries and cylinders, and extended it to the torus.

This formalism allows to construct a relatively simple unifying picture of a series of amusing properties of $\mathbb{Z}_2$ topologically ordered states enriched by translational symmetry,~\cite{Wen2009,Wen2010,Wen2003,Kou2008,Cho2012} including their anomaluous GSD dependence on the size of the torus and the appearance of dangling Majorana modes at the boundaries of open lattices even in states whose bulk topological order is identical to the Toric Code. This formalism has also allowed us to unravel the intimate connection between such anomalies of $\mathbb{Z}_2$ topological ordered states and the phenomenon of `weak symmetry breaking'.~\cite{Kitaev2006} Weak symmetry breaking is a remarkable phenomenon in which the vacuum of a phase of matter remains invariant under a symmetry of the Hamiltonian, but the symmetry is in a sense broken by its quasi-particles. This is only possible if the quasi-particles are non-local anyons, and more precisely, it is the phenomenon in which the symmetry action on certain anyons switches them into a distinct anyon type belonging to a different super-selection sector, and therefore, cannot be implemented by any local physical operation.

These phenomena in translationally invariant $\mathbb{Z}_2$ topologically ordered states are intimately related to the topological classification of translational invariant  BdG Hamiltonians.~\cite{Wen2009,Wen2010,Sato2010,Sato12010} Such 2D fermionic paired states with translational symmetry (Class D plus lattice translations) can be classified by their Chern number and three other $\mathbb{Z}_2$ topological parity indices (also known as Pfaffian indicators), namely, each phase can be labeled by a vector $(C,\zeta_{K_x},\zeta_{K_y},\zeta_{\text{AI}})$, where $C\in \mathbb{Z}$ and $\zeta= \{0,1\}$. These indices have a natural physical interpretation: $C$ is the well known Chern number `strong' index counting the chirality of edge Majorana modes, and all the $\zeta$ indices are `weak' indices accounting if the phase contains stacks of lower dimensional topological superconducting phases. $\zeta_{K_x} (\zeta_{K_y})=1$ corresponds to having a stack of Kitaev wires oriented in the $x$- ($y$-) direction, and $\zeta_\text{AI}=1$ corresponds to having a filled `Atomic Insulator' band with one fermion per unit cell. We have also provided an argument for why all of these `weak' topological indices are robust against fermion interactions and self-averaging disorder that respect translational symmetry. Although in the literature of BdG Hamiltonians the phases with $\zeta_\text{AI}=1$ are often viewed as trivial, for our purposes it is crucial to keep track of this index, since in the case of $\mathbb{Z}_2$ topologically ordered states on a torus, one must discard states with an odd number of fermions as unphysical, thus leading to anomalous GSD dependence on the size of the torus for states with $\zeta_\text{AI}=1$. More generally, whenever at least one of the indices $(C,\zeta_{K_x},\zeta_{K_y},\zeta_{\text{AI}})\in (\mathbb{Z},\mathbb{Z}_2,\mathbb{Z}_2,\mathbb{Z}_2)$ is odd, the system will have a GSD that is not $4$ in certain tori, and this can occur even when $C=0$ in spite of the bulk topological properties of its anyons remaining the same of the Toric Code. Eq.~(\ref{Eqn:AnomalousGSD}) provides a general formula to compute the GSD in any system size for these states.

The phenomenon of weak breaking of translational symmetry in $\mathbb{Z}_2$ topologically ordered states in two dimensions occurs when the $\varepsilon$-fermions form a paired state which contains a stack of Kitaev wires, namely, when either of the indices $(\zeta_{K_x},\zeta_{K_y})$ is non-zero. Moreover, since these states are made from stacks of Kitaev wires they feature dangling Majorana modes that will generally hybridize into a 1D boundary gapless Majorana spectrum protected by translational symmetry in the edge. The reason such phases display weak symmetry breaking of translations in the bulk stems from the fact that even though the $e$-particles are dynamically frozen, the operator that transports them to neighboring vertices needs to be modified in the presence of the non-trivial background state of the $\varepsilon$-fermions. Specifically, because the $e$-particle is viewed as a source $\pi$-flux by the $\varepsilon$-fermions, it carries a `string' that twists the sign of fermion hopping. Therefore, when the $\pi$-flux hops across a Kitaev wire, it effectively flips its boundary conditions from periodic to anti-periodic (or vice-versa depending on the original boundary condition of the wire). Since the fermion parity of a Kitaev wire in its non-trivial phase depends on the boundary conditions twists, such hop will necessarily create a Bogoliubov fermion and therefore cannot be a symmetry as it would change the energy of the state. More importantly there is no way to restore the system back into its ground state in any local manner because it will require the destruction of a single $\varepsilon$-fermion. Therefore the $e$-particles cannot hop locally to any neighboring vertex if such hop requires crossing an odd number of $\varepsilon$-fermion Kitaev wires. However, when the $e$-particle hops to a second neighbor vertex by crossing two Kitaev wires, the ground state can be restored by an inter-wire $\varepsilon$-pair creation operator, which is local. As a result the $e$-particles break into two distinct superselection sectors residing in two sublattices of vertices with non-trivial mutual semionic statistics when the $\varepsilon$-fermions form a stack of Kitaev wires.

One of the advantages of employing the exact $\mathbb{Z}_2$ flux-attachment description is that it provides an exact one-to-one rewriting of the physical states of the Hamiltonian without the need to locally enlarge the Hilbert space as it is often done in parton descriptions. More precisely, the $\mathbb{Z}_2$ flux-attachment only has global unphysical parity symmetries in the torus, but no unphysical symmetries in the fully open lattice or in the cylinder. In practice the unphysical parity symmetries in the torus can be dealt with easily, by simply restricting to states with an even number of $\varepsilon$-fermions and even number of $e$-particles. Using this construction we have written down a model that interpolates from the Toric Code~\cite{Kitaev2003} to the Kitaev honeycomb model,~\cite{Kitaev2006} and that realizes a variety of the non-trivial phases described above. In addition, our extension of this technique to the open and cylindrical lattices allowed us to compute explicitly their edge spectrum even away from the ideal fixed point commuting projector Hamiltonians. This is ultimately possible thanks to the local symmetry (gauge structure) that freezes the motion of isolated $e$- and $m$-particles and only allows fluctuations of the $\varepsilon$-fermions, thus providing a machinery allowing to construct exactly solvable models for any free fermion Hamiltonian. Although we have focused only on enforcing lattice translational symmetry, this machinery is naturally suited to study the interplay of $\mathbb{Z}_2$ topological order and symmetry in many other cases.

\section*{Acknowledgement}
We are grateful to Vijay Shenoy, \'Oscar Pozo Oca\~na, Chuan Chen, Joseph Maciejko, Max Geier, Piet Brouwer, Luka Trifunovic, Seishiro Ono, Masatoshi Sato, Sergej Moroz, Umberto Borla and Cesar Gallegos for useful discussions and correspondence.

\appendix
\begin{figure*}
	
	\includegraphics*[width=0.5\linewidth]{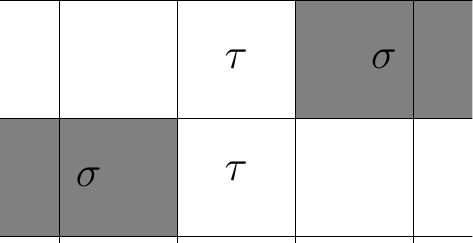}\llap{
		\parbox[b]{17cm}{\large (a)\\\rule{0ex}{4.7cm}
	}}~~
	\includegraphics*[width=0.3\linewidth]{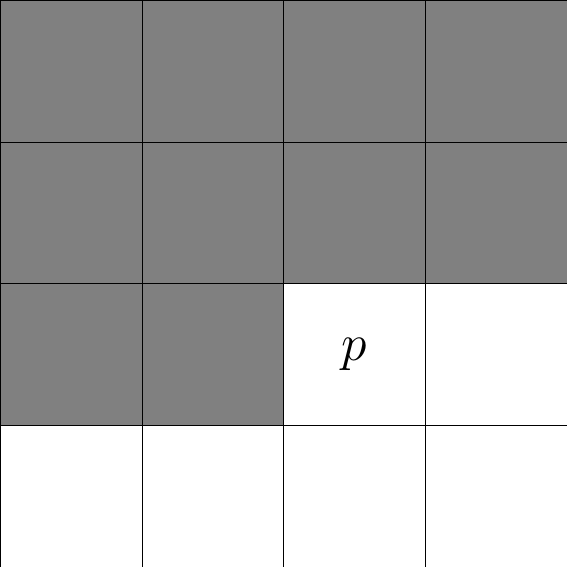}\llap{
		\parbox[b]{10cm}{\large (b)\\\rule{0ex}{4.7cm}
	}}
	
	\caption{\label{Fig:JordanWigner} (a) Visual representation of $U_y$ in Eq.~(\ref{Eqn:DualMapping3}) in the dual space. Shaded plaquettes indicate those that enter into the products in Eq.~(\ref{Eqn:DualMapping3}). (b) Visual representation of `stringing' of the lattice in Eq.~(\ref{Eqn:MajoranaFermions2}) for given $\gamma, \gamma'$ on plaquette $p$. The shaded plaquettes indicate those that enter into the products in Eq.~(\ref{Eqn:MajoranaFermions2}). }
\end{figure*}

\section{2D Jordan-Wigner Transformation of Majorana Fermions in Eq.~(\ref{Eqn:OperatorDefinitions})} \label{Appendix:Duality}

In this Section, we provide an alternative derivation of the fermion mapping Eq.~(\ref{Eqn:OperatorDefinitions}), which takes into account directly the dependence of boundary condition on topological operators discussed in Section~\ref{Sec:TorusGeometry}. In this construction, we will introduce an intermediate dual Hilbert space with bosonic degrees of freedom. And subsequently, we will map these dual bosonic degrees of freedom into the fermionic ones that are discussed throughout the main text, via a mapping that resembles the more conventional Jordan-Wigner transformation. For simplicity we will restrict to the subspace containing no $e$-particles and the torus geometry.

We take the dual bosonic degrees of freedom to reside in the plaquettes and their occupation to coincide with that of the $\varepsilon$-fermion occupation number. Namely, if we denote by $N=0, 1$ the local occupation of the dual bosons at a given plaquette, by $\sigma$ the dual boson parity and by $\tau$ the $x$-like Pauli matrix that swaps the boson parity, we have: 
\be\label{Eqn:DualSpace}
\sigma\ket{N}=(-1)^N \ket{N},~\sigma\tau=- \tau \sigma.
\ee
In this appendix we will restrict our discussion here to the representation of the susbpace physical Hilbert space in which the $e$-particle configuration has been fixed with $\Gamma^e_v=1$ from Eq.~(\ref{Eqn:eDefinition}). Thus the task is to find a representation of the elementary operators that commute with $\Gamma^e_v$, namely, $\Gamma^\varepsilon_p$ and $U_{x,y}$ in Eqs.~(\ref{Eqn:FermionParity}) and (\ref{Eqn:PairCreation}) respectively.

As the next step, we express these operators in the intermediate Hilbert space in terms of $\tau$ and $\sigma$ defined in Eq.~(\ref{Eqn:DualSpace}). The mappings for these operators must satisfy the same commutation relations as those in the underlying spin Hilbert space. First, the fermion parity operators $\Gamma^\varepsilon_p$ in Eq.~(\ref{Eqn:FermionParity}) is mapped by definition into:
\be\label{Eqn:DualMapping}
\Gamma^\varepsilon_p \rightarrow \sigma_p.
\ee
The global constraint in Eq.~(\ref{Eqn:Constraints2}) then becomes:
\be\label{Eqn:Constraints3}
\prod_{p\in \text{lattice}} \sigma_p=1.
\ee
The transport operator $U_{x,p}$ in Eq.~(\ref{Eqn:PairCreation}) creates a pair of fermions on plaquette $p$ and the plaquette to its right. Thus it anti-commutes with $\Gamma^\varepsilon$ on these two plaquettes. To satisfy the correct commutation relations, we choose the second duality mapping to be:
\be \label{Eqn:DualMapping1}
U_{x,p} \rightarrow \tau_{n,k}\tau_{n,k+1},
\ee
where $p=(n,k)$, and $n$ designates rows and $k$ columns.

So far it seems that the mappings above are identical with the usual bosonic duality in $\mathbb{Z}_2$ lattice gauge theories.~\cite{Oscar2020} The difference arises for the mapping of vertical translation operators $U_{y,p}$ in Eq.~(\ref{Eqn:PairCreation}). This is because $U_{y,p}$ anti-commutes with $U_{x}$ on plaquettes to its North and East, whereas in the bosonic duality all $U_x$ and $U_y$ commute. Therefore, $U_y$ cannot be simply mapped into $\tau_i \tau_j$ but must contain additional terms. Here we choose for later convenience:
\be \label{Eqn:DualMapping3}
U_y \rightarrow \tau_{n,k} \tau_{n+1,k} \bigg(\prod_{i<k} \sigma_{n,i}\bigg)\bigg(\prod_{i>k} \sigma_{n+1,i}\bigg).
\ee 
The product of $\sigma$ is taken over all plaquettes to the left of $(n,k)$ and to the right of $(n+1,k)$, and is visually represented in Fig.~\ref{Fig:JordanWigner}. From this expression, one can verify that these operators satisfy the same commutation relations as those defined in terms of the underlying spins in Eq.~(\ref{Eqn:OperatorDefinitions}).

So far we have not taken into account that the Torus geometry imposes a global parity constraints for $\Gamma^e_v$ and $\Gamma^\varepsilon$ in Eqs.~(\ref{Eqn:eConstraint}) and (\ref{Eqn:Constraints2}). The parity constraint in Eq.~(\ref{Eqn:Constraints3}) reduces the intermediate dual Hilbert space dimension to $2^{L_xL_y-1}$, but the underlying spin Hilbert still has $2^{L_xL_y+1}$ degrees of freedom after specifying the eigenvalues of all $L_xL_y-1$ independent $\Gamma^e_v$ operators (see a similar argument at the beginning of Section~\ref{Sec:TorusGeometry}). This apparent mismatch of dimensionality originates from the fact that we have not yet accounted for the four topological degrees of freedom associated with $T_x$ and $T_y$ in Eq.~(\ref{Eqn:t'Hooft1}). This can also be seen from the fact that $T_{x,y}$ are related to $U_{x,y}$ and $\Gamma^\varepsilon$ operators by Eq.~(\ref{Eqn:Consistency}), which would contradict, for example, the identity from Eq.~(\ref{Eqn:DualMapping1}) that $\prod U_x$ taken over a row would be unity in the intermediate Hilbert space. As we shall show below, this can be resolved by a small modification of the bosonic duality mappings for hopping across certain `branch-cuts' of the lattice, as discussed in the main text surrounding Eq.~(\ref{Eqn:BoundaryHopping}) (see also Fig.~\ref{Fig:BranchCut}) and also further discussed in Ref.~\onlinecite{Oscar2020}.

Therefore, in order to be able to represent the different possible values of $\{T_x,T_y\}$, we introduce additional dual $\mathbb{Z}_2$ valued operators $\theta_n$, each one associated with the $n$-th row of the lattice. These operators allow to represent the horizontal hopping operator associated with crossing the vertical branch-cut (see Fig.~\ref{Fig:BranchCut}) by modifying the last horizontal hopping of each row:
\be\label{Eqn:DualMapping2}
U_{x} \rightarrow \tau_{n,L_x}\tau_{n,1}\vartheta_n.
\ee
$\vartheta_n$ is chosen to commute with themselves and with all other dual operators. Then we have:
\be\label{Eqn:RowHopping}
\prod_{\text{row}~n} U_x \rightarrow \vartheta_n. 
\ee
Up to this point we have enlarged the dual Hilbert space by a large number of states, because we have introduced one $\theta_n$ for every row. However these operators are not independent. To see this, we rewrite Eq.~(\ref{Eqn:t'Hooft1}) as:
\be
T_x = - \prod_{p\in \text{row}~n} U_{x,p} \prod_{p\in \text{row}~n} \Gamma^\varepsilon_p,
\ee
which gives another constraint for dual bosonic operators:
\be\label{Eqn:t'Hooft2}
T_x \rightarrow -\vartheta_n\Pi_n ,~\Pi_n = \prod_{p\in \text{row}~n} \sigma_p.
\ee
Eq.~(\ref{Eqn:t'Hooft2}) holds for each row separately. However, since $T_x$ is row-independent, $\vartheta_n$ is related to $\vartheta$ on other rows by:
\be\label{Eqn:t'Hooft3}
\vartheta_n \Pi_n = \vartheta_m \Pi_m.
\ee
From Eq.~(\ref{Eqn:t'Hooft3}) one can see that only one of the $\vartheta_n$ operators is independent. Therefore, $T_x$ is taken into account by introducing $\vartheta_n$ without any further enlargement of the Hilbert space..

For vertical hopping, we introduce a $\mathbb{Z}_2$ operator $\varphi$ for the last vertical hopping on each column analogous to $\vartheta_n$:
\be\label{Eqn:DualMapping4}
U_{y} \rightarrow \tau_{L_y,k}\tau_{1,k}\bigg(\prod_{i<k} \sigma_{L_y,i}\bigg)\bigg(\prod_{i>k} \sigma_{1,i}\bigg) \varphi_k.
\ee
Multiplying $U_y$ across a column using Eqs.~(\ref{Eqn:DualMapping3}) and (\ref{Eqn:DualMapping4}) and substituting Eq.~(\ref{Eqn:Constraints3}) gives:
\be\label{Eqn:t'Hooft5}
T_y = - \prod_{p\in \text{column}~k} U_{y,p} \prod_{p\in \text{column}~k} \Gamma^\varepsilon_p = -\varphi_k.
\ee
Eq.~(\ref{Eqn:t'Hooft5}) then relates $T_y$ to $\varphi_k$. By a similar reasoning to that above, we can see that there is only one independent $\varphi_k$ and therefore, a one-to-one correspondence with values of $T_y$.

Eq.~(\ref{Eqn:DualMapping3}) has the advantage of admitting a definition of Majorana fermions in Eq.~(\ref{Eqn:OperatorDefinitions}) as a natural extension of Jordan-Wigner transformation in 1D. Similarly to the 1D case, $\gamma, \gamma'$ are non-local and contain a `string' of $\sigma$ operators in the following way: on a given plaquette $(n,k)$, the string goes through all rows above row $n$ from left to right and, on row $n$, goes to the column $k$ from the left; see Fig.~\ref{Fig:JordanWigner}. Explicit definitions are
\bea\label{Eqn:MajoranaFermions2}
\begin{split}
	\gamma_{n,k} &= i\bigg(\prod_{i>n,j} \sigma_{i,j} \bigg)\bigg(\prod_{i< k}  \sigma_{n,i} \bigg)\sigma_{n,k}\tau_{n,k}, \\
	\gamma^\prime_{n,k} &=\bigg(\prod_{i>n,j} \sigma_{i,j} \bigg)\bigg(\prod_{i< k}  \sigma_{n,i} \bigg)\tau_{n,k}.
\end{split}
\eea
$\gamma$ and $\gamma'$ satisfy the fermion anti-commutation relations and, substituting Eq.~(\ref{Eqn:MajoranaFermions2}) into Eqs.~(\ref{Eqn:DualMapping1}) and (\ref{Eqn:DualMapping3}), we recover Eq.~(\ref{Eqn:OperatorDefinitions}) first obtained in Ref.~\onlinecite{Chen2018}.

Eq.~(\ref{Eqn:MajoranaFermions2}) also gives directly the relation between $\{T_x,T_y\}$ and the fermion boundary conditions along $x$- and $y$-directions. In the fermionic Hilbert space, periodic and anti-periodic boundary conditions can be represented by an additional $\pm1$ in Majorana fermion hopping across the lattice `branch-cut':
\be
U_p \rightarrow \pm i \gamma \gamma'.
\ee
See Eq.~(\ref{Eqn:BoundaryHopping}). Using Eq.~(\ref{Eqn:MajoranaFermions2}), hopping across the `branch-cut' along the $x$-axis gives:
\be
\pm i\gamma_{n,L_x}\gamma'_{n,1} = \mp \tau_{n,L_x}\tau_{n,1} \prod_k \sigma_{n,k}.
\ee
Comparing with Eqs.~(\ref{Eqn:DualMapping2}) and (\ref{Eqn:t'Hooft2}), we obtain $\vartheta_n=\mp \Pi_n$ and $T_x=\pm 1$ for periodic and anti-periodic boundary conditions respectively. Similarly, for hopping across the `branch-cut' along the $y$-axis:
\be\label{Eqn:ColumnHopping}
\pm i\gamma_{1,k}\gamma'_{L_y,k} = \mp \tau_{L_y,k}\tau_{1,k}\bigg(\prod_{i<k} \sigma_{L_y,i}\bigg)\bigg(\prod_{i>k} \sigma_{1,i}\bigg),
\ee
where we have used Eq.~(\ref{Eqn:Constraints3}). Comparing Eq.~(\ref{Eqn:ColumnHopping}) with Eqs.~(\ref{Eqn:DualMapping4}) and (\ref{Eqn:t'Hooft5}), we see that $T_y=-\varphi_k = \pm 1$ for periodic and anti-periodic boundary conditions. Thus we have re-derived the relation between $T_{x,y}$ and the corresponding fermion boundary conditions, which is obtained in Section~\ref{Sec:TorusGeometry} using another method.

Finally, as a consistency check, we show that the dual bosonic mappings in Eqs.~(\ref{Eqn:DualMapping}), (\ref{Eqn:DualMapping1}), (\ref{Eqn:DualMapping3}) and (\ref{Eqn:DualMapping2}) reproduce the ground state degeneracy of the model considered in Section~\ref{Sec:Model}. As an example, we study Eq.~(\ref{Eqn:Decoupled}) in the intermediate Hilbert space at $h_z=h_y=0, h_x>0$, and compare with results in Section~\ref{Sec:Torus}. We first consider the Torus. Multiplying Eq.~(\ref{Eqn:t'Hooft2}) over all rows and using Eq.~(\ref{Eqn:Constraints3}) gives:
\be\label{Eqn:t'Hooft4}
(-T_x)^{L_y} = \prod_{n\in \text{all rows}} \vartheta_n.
\ee 
In the ground state, $U_{x}=1$ and $\vartheta_n=1$. Eq.~(\ref{Eqn:t'Hooft4}) gives:
\be
(-T_x)^{L_y} = 1.
\ee
For even $L_y$, this relation is trivial and both $T_x$ values are allowed: the ground state is $4$-fold degeneracy labeled by $T_x, T_y$. For odd $L_y$, Eq.~(\ref{Eqn:t'Hooft4}) forbids $T_x=1$ (periodic boundary condition) and Eq.~(\ref{Eqn:t'Hooft2}) gives $\Pi_n=1$ on each row. The ground state is thus only $2$-fold degenerate labeled by $T_y$. For $h_x<0$, $U_x=-1$ and Eq.~(\ref{Eqn:t'Hooft4}) becomes:
\be
(-T_x)^{L_y} = (-1)^{L_xL_y}.
\ee
So for $L_y$ even, both $T_x$ values are allowed and for $L_y$ odd, $T_x = \pm 1$ for $L_x$ odd or even. The dependence of GSD on $L_y$ is the same as for $h_x>0$, which agrees with Section~\ref{Sec:Torus}. 

In the case of open boundary along $y$-axis only, Eq.~(\ref{Eqn:t'Hooft2}) still holds, so that parities on each chain are still related. But without the parity constraint in Eq.~(\ref{Eqn:Constraints3}) there is no restriction on the value of $\Pi_n$ hence $T_x$, and the difference between $L_y$ odd and even disappears.

\section{Independence of Parity operators in TC in an open Lattice}\label{Appendix:TCOperatorIndependence}

In this Section we prove that, in an open lattice, there are no global constraints for $G^e_v$ and $G^m_p$, whereas on a Torus constraints in Eq.~(\ref{Eqn:Constraints1}) exists.

Global constraints for $\mathbb{Z}_2$ operators $G^e_v$ and $G^m_p$ can be written in the following form:
\be\label{Eqn:TCOpenLatticeConstraint}
G^e_{v_1} = F_1(G^e_v,G^m_p),~G^m_{p_1} = F_2(G^e_v,G^m_p),
\ee
where $v_1$ and $p_1$ are given vertex and plaquette in the lattice. $F_1$ and $F_2$ are functions of $G^e_v$ and $G^m_p$ on all other vertices and plaquettes. If there is an operator that anti-commutes with $G^e_{v_1}$ or $G^m_{p_1}$ but commutes with all other $G^e_v$ and $G^m_p$, then it would contradict the existence of constraint (\ref{Eqn:TCOpenLatticeConstraint}) and global constraints cannot exist. However, such operators are just single-particle creation operators for $e$ and $m$ particles introduced in Section~\ref{Sec:OpenLattice}. Therefore the existence of single particle creation operators that commute with all other parities and only anti-commute with the parity of the plaquette or vertex of interest, implies that a constraint such as that in Eq.~(\ref{Eqn:TCOpenLatticeConstraint}) cannot exist in open lattices or cylinders.

\section{Fermion Creation operators in Open and Cylindrical Lattices}\label{Appendix:FermionCreation}
In open and cylindrical lattices with finite size, the fermionic even parity constraint (\ref{Eqn:Constraints2}) no longer holds, and single $\varepsilon$-particles can be created. In this Section, we describe how to construct single Majorana fermion operators on these lattices in the underlying spin Hilbert space.

\begin{figure}
	
	\includegraphics*[width=0.8\linewidth]{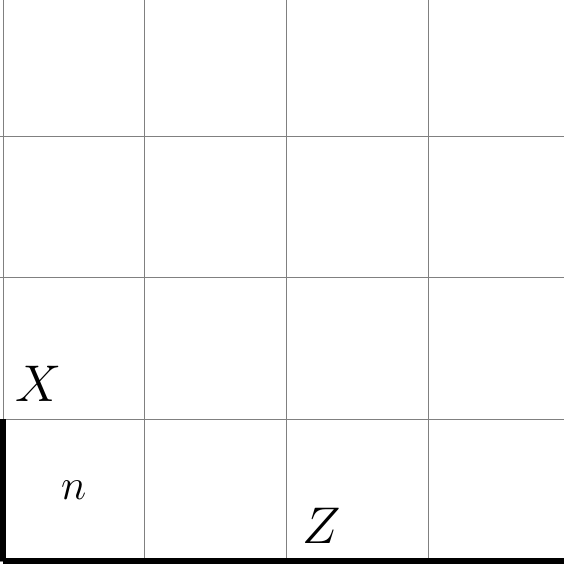}~~
	\caption{\label{Fig:OpenBoundary} Majorana fermion creation operator $\gamma_n'$ in an open lattice. The lattice size is $L_x=L_y=4$.}
	
\end{figure}

We first consider the open lattice. It is sufficient that the operator is found for a single Majorana fermion on a given site, since other Majorana operators can be obtained by multiplying it with parity and pair-creation operators in Eq.~(\ref{Eqn:OperatorDefinitions}). Such an operator for the bottom left plaquette $n$ of the open lattice is shown in Fig.~\ref{Fig:OpenBoundary} as a product of a single $X$ and a $Z$ line along bottom of the lattice. The operator has the following physical meaning: because of open boundaries, the $X$ creates a single $m$-particle at plaquette $n$ while the $Z$ line creates an $e$-particle on the south-east edge, which is then transported along the lower edge to the SW vertex of $n$ and forms an $\varepsilon$-particle. This operator is mapped into $\gamma'_n$ in the fermion Hilbert space: it anti-commutes with $\Gamma^\varepsilon_n, U_{y,n}$ but commutes with all other local fermion operators and $\Gamma^e_v$ and, from the mapping (\ref{Eqn:OperatorDefinitions}), it follows that it creates a single $\gamma'$ on the plaquette $n$. We note that $\gamma$ or $\gamma'$ operators can be constructed similarly for all plaquettes along west and south edges, which differ from the one above by appropriate product of $U_{x,y}, \Gamma^\varepsilon_p$ and $\Gamma^e_v$. In fact, it can be shown that the string operator in Fig.~\ref{Fig:OpenBoundary} is mapped to $\tau_n$ in the notation of Sec.~\ref{Appendix:Duality}, and $\gamma, \gamma'$ thus defined correspond to a different Jordan-Wigner convention than in Sec.~\ref{Appendix:Duality}.

On a cylinder, an analogous operator can be defined as the same product of $X$ on the west edge and a $Z$ line which transports a single $e$ across the lattice. However, due to the periodicity along $y$-direction, the $Z$ line anti-commutes with $U_y$ along the path, in addition to anti-commuting with $\Gamma^\varepsilon_n, U_{y,n}$. This non-locality is a result of the dependence of boundary conditions along $y$-direction on particle configurations in the lattice, as given by Eq.~(\ref{Eqn:CylinderTwist}). Creating a single fermion corresponds to a change of boundary conditions which changes the sign of $U_y$ along a horizontal branch cut given by the $Z$ line above. Thus, the operator swaps boundary conditions. We emphasize that these operators do not map into single Majorana fermions, since they clearly commute between themselves.

\section{Fixed Point Model with a diagonal-stacking of Majorana fermions}\label{Appendix:DiagonalStacking}
\begin{figure}	
	\includegraphics*[width=0.8\linewidth]{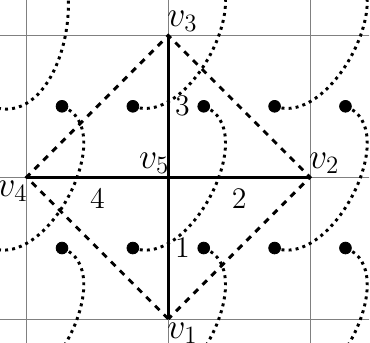}	
	\caption{\label{Fig:WenPhase} Couplings between Majorana modes $\gamma', \gamma$ in Eq.~(\ref{Eqn:WenHamiltonian1}) are illustrated visually as dotted lines in phases KW$_{x+y,\zeta}$. The $e$-particles residing in vertices belonging to two sub-lattices acquire mutual semionic statistics. They become $e'$ ($v_{1-4}$) or $m'$ ($v_5$) along the  diagonal of the square unit cell but has anyonic statistics with respect to its nearest neighbor. }
\end{figure}

As mentioned in Section~\ref{Sec:TopologicalSuperconductor}, KW$_{x+y,\zeta}$ denotes the class of states that are topologically equivalent to stacking 1D Kitaev wires along the diagonal direction. The ideal or fixed point Hamiltonian associated with this phase can be realized by choosing the corresponding $C_p$ in Table~\ref{Table:IdealOperators} for the ideal Hamiltonian~\ref{Eqn:IdealHamiltonian}:
\be\label{Eqn:WenHamiltonian}
H = -\Delta_e \sum_v \Gamma^e_v -\Delta_\varepsilon \sum_p   \Gamma^\varepsilon_{NE(p)} U_{y,E(p)}U_{x,p},
\ee
where $E(p)$ and $NE(p)$ are plaquettes to the east and north-east of plaquette $p$. Substituting Eq.~(\ref{Eqn:OperatorDefinitions}), Eq.~(\ref{Eqn:WenHamiltonian}) is mapped into the following fermionic Hamiltonian:
\be\label{Eqn:WenHamiltonian1}
H = -i \Delta_\varepsilon \sum \gamma_i \gamma'_k.
\ee
The pairing of Majorana modes is depicted by curved dotted lines in Fig.~\ref{Fig:WenPhase}. The BdG spectrum for Eq.~(\ref{Eqn:WenHamiltonian1}) has the same form as Eq.~(\ref{Eqn:BCSHamiltonian}) with 
\be
\begin{split}
	\varepsilon(\bm{k}) &= -2\Delta_\varepsilon \cos (k_x+k_y),\\
	\Delta(\bm{k}) &= -2i \Delta_\varepsilon\sin (k_x+k_y).
\end{split}
\ee
Parity topological matrices $\zeta_{ij}$ for $\Delta_\varepsilon >0 (<0)$ coincides with KW$_{x+y,0}$ phase (KW$_{x+y,1}$ phase).

We now show that there is `weak breaking' of translational symmetry in these phases: the $e$-particles split into two sectors of effective anyons $e'$ and $m'$ in the ground state, and lattice translations along both directions permute them. For example, in Fig.~\ref{Fig:WenPhase}, when $e$ resides in vertices $v_{1-4}$ become $e'$ while on $v_5$ it becomes $m'$. This is similar to the Wen's plaquette model.~\cite{Wen2003} To show this, we proceed analogously to Section~\ref{Sec:WeakSymmetry}, by finding the modified $e$-translation operators in the ground state subspace of KW$_{x+y,\zeta}$. Such an operator must commute with the corresponding Majorana bilinear terms in Eq.~(\ref{Eqn:WenHamiltonian}). They can be found only for translations between diagonals of a square. For example, in Fig.~\ref{Fig:WenPhase}, translations between $v_1, v_2$ and $v_2,v_3$ are
\be\label{Eqn:AnyonTransport1}
T^e_{v_2v_1} = Z_1 Z_2,~T^e_{v_3v_2} = i (Z_2 Z_3) \times (X_3 Z_4).
\ee
The factor $i$ imposes $(T^e_{v_3v_2})^\dagger = T^e_{v_3v_2}$. Eq.~(\ref{Eqn:AnyonTransport1}) can be understood intuitively similarly to the horizontal stacking case in Section~\ref{Sec:WeakSymmetry}. We draw Majorna pairings with curved lines in the form in Fig.~\ref{Fig:WenPhase}. When $e$ is transported from $v_1$ through $v_5$ to $v_2$, it cuts through the same Majorana bond twice, therefore does not change the fermion parity associated with such a pair of Majorana modes. However, going from $v_2$ through $v_5$ to $v_3$,it cuts through two different bonds, annihilating two Majorana fermions, which are then created by the pair creation operator $X_3 Z_4$. The loop translation operator on the ground state $|0\rangle$ along the dashed line in Fig.~\ref{Fig:WenPhase} now gives:
\be
T^e_{v_4v_1}T^e_{v_3v_4}T^e_{v_2v_3}T^e_{v_1v_2}|0\rangle = \Gamma^e_{v_5} |0\rangle,
\ee
where we used the ground state identity $\pm \Gamma^\varepsilon_{NE(p)} U_{y,E(p)}U_{x,p}|0\rangle =|0\rangle$ for KW$_{x,0}$ and KW$_{x,1}$ respectively. This demonstrates how the $e$-particles in one sublattice picks up a $-1$ sign when they are transported in a loop that encloses an odd number of $e$-particles in the other sublattice.

\section{Recasting the KW$_{x,y}$ as an ordinary Toric Code}\label{Appendix:ToricCodeGS}

In this Section we show that, the ideal fixed point Hamiltonians associated with the KW$_{x,\zeta}$ (KW$_{y,\zeta}$) phases described in Section~\ref{Sec:TopologicalSuperconductor} can be recast as an ordinary TC in an infinite lattice or for $L_y$ ($L_x$) even. 

\begin{figure}	
	\includegraphics*[width=0.7\linewidth]{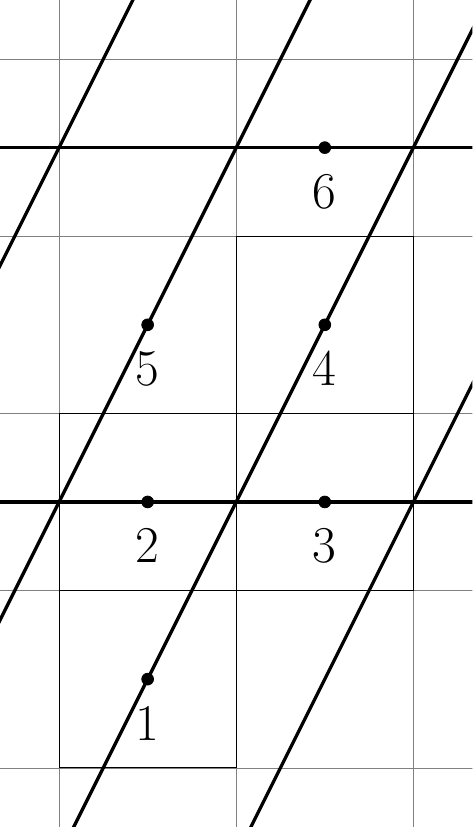}
	
	\caption{\label{Fig:ToricCodeGS} New lattice in the subspace of $U_{x}=\pm1$ by treating the right and lower links of a given plaquette as one degree of freedom at the center. $\Gamma^e_v$ becomes on odd rows $G^{e'}_v$ and $G^{m'}_p$ on even rows in the TC.}
\end{figure} 

For this purpose, we first consider the ideal fixed point Hamiltonian for phase KW$_{x,\zeta}$, which corresponds to $h_z=\delta=0$ in Eq.~(\ref{Eqn:Decoupled}), and project it into the following subspace satisfied by the ground states of KW$_{x,1}$ and KW$_{x,0}$ respectively:
\be\label{Eqn:GroundStateInequality}
U_{x,p}=\mp1,~h_x \lessgtr 0, 
\ee
To demonstrate our statement at the beginning of this Section, we will show that the ideal fixed point Hamiltonian projecting to the subspace of Eq.~(\ref{Eqn:GroundStateInequality}) becomes the TC for $L_y$ even. Physically, Eq.~(\ref{Eqn:GroundStateInequality}) corresponds to the limit:
\be\label{Eqn:GroundStateInequality1}
0<\Delta_e \ll |h_x|.
\ee
The inequality (\ref{Eqn:GroundStateInequality1}) means a large superconducting gap for $\varepsilon$-fermions, and the low-energy subspace of Eq.~(\ref{Eqn:GroundStateInequality}) has no $\varepsilon$-particles. However, as we shall see, the $e$-particles on top of this non-trivial superconducting vacuum are split into two groups of anyons ($e'$ and $m'$) which can be identified with those of TC, while $\Gamma^e_v$ becomes accordingly either $G^{e'}$ or $G^{m'}$ defined in Eq.~(\ref{Eqn:TCOperators}).

In the subspace given by Eq.~(\ref{Eqn:GroundStateInequality}), we can choose a new basis such that the horizontal and vertical links in $U_{x,p}$ are simultaneously diagonal with respect to $Z$ and $X$. For $h_x\lessgtr 0$, $Z$ and $X$ have opposite (the same) eigenvalues, and can be treated as one degree of freedom defined on the plaquette $p$. Thus, in the notation of Fig.~\ref{Fig:OperatorDefinitions}, we have the mapping:
\be
Z_3\rightarrow \sigma^z_p, X_5 \rightarrow \mp \sigma^z_p,~h_x\lessgtr 0.
\ee 
$\sigma^z_p$ is the third Pauli matrix acting on the plaquette $p$ with the eigenvalue of $Z_3$. A simultaneous operation of $X_3$ and $Z_5$ on horizontal and vertical links of $p$ anticommutes with $Z_3$ or $X_5$ yet commutes with $U_{x,p}$, so the ground state identity Eq.~(\ref{Eqn:GroundStateInequality}) is still satisfied. As a result we have another mapping within the subspace:
\be
X_3Z_5 \rightarrow \sigma^x_p.
\ee
$\Gamma^e_v$ then becomes a four plaquette operator $\Gamma^e_p$ defined on the plaquette to the north-east of $v$. For $h_x\lessgtr0$ (both KW$_{x,0}$ and KW$_{x,1}$ phases), $\Gamma^e_p$ has the same form. For example, in Fig.~\ref{Fig:ToricCodeGS}:
\be\label{Eqn:EffectiveAnyons}
\Gamma^e_3 \rightarrow \sigma^z_1 \sigma^y_2 \sigma^y_3 \sigma^z_4.
\ee
To make more explicit the connection with the ordinary TC, it is more convenient to define a new lattice in which Eq.~(\ref{Eqn:EffectiveAnyons}) have the explicit form of $G^{e'}$ and $G^{m'}$ operators in TC. We join, for all $\Gamma^e_p$ on odd rows, the centers of plaquettes and treating them as the mid-points of links of the new lattice. For example, in Fig.~\ref{Fig:ToricCodeGS} this is done for $\Gamma^e_3$ by joining plaquettes $2, 3$ and $1, 4$. Then $\Gamma^e_p$ becomes the $G^{e'}$ and $G^{m'}$ operators in TC on odd rows and even rows  ($\Gamma^\varepsilon_3$ and $\Gamma^\varepsilon_4$ in Fig.~\ref{Fig:ToricCodeGS}). For an infinite lattice or $L_y$ even, the separation of $e$ into $e'$ and $m'$ is consistent. However, for $L_y$ odd this construction breaks down, as can be seen by the following. We designate $e$ from the first row as $e'$ and then $e$ from the second row as $m'$ and so on. Repeating this procedure through the entire lattice, we see that, upon returning to the first row from the $L_y$-th row, the $e$ on the first row should become $m'$ instead of $e'$ in contradiction to the initial designation. 

Thus the Hamiltonian Eq.~(\ref{Eqn:Decoupled}) in the subspace given by Eq.~(\ref{Eqn:GroundStateInequality}) is equivalent to the TC at low energies for an infinite lattice or $L_y$ is even. This means that, for a finite lattice with $L_y$ even, the usual Toric Code constraints apply:
\be\label{Eqn:EffectiveTCConstraints}
\prod_{p\in \text{odd rows}}\Gamma^{e}_p =1, \prod_{p\in \text{even rows}}\Gamma^{e}_p =1.
\ee
In particular, the ground state of the ideal fixed point Hamiltonian of KW$_{x}$ phases is always in the subspace of Eq.~(\ref{Eqn:GroundStateInequality}). Since, as discussed above, for odd $L_y$, only the constraint in Eq.~(\ref{Eqn:Constraints1}) applies, the ground state is $2$-fold degenerate for $L_y$ odd and $4$-fold degenerate for $L_y$ even, which agrees with the conclusion in Section~\ref{Sec:Model}. Above results can also be obtained by counting degrees of freedom of the Hamiltonian in Eq.~(\ref{Eqn:Decoupled}). The derivation is given in Section~\ref{Appendix:GSDSpinLattice}. 

We note that a translation along the $y$-direction by unity in the original lattice exchanges $e'$- and $m'$-particles in the new lattice. This is another manifestation of the `weak symmetry breaking' mentioned in Section~\ref{Sec:WeakSymmetry}. 

Above considerations can be extended to KW$_{y,\zeta}$ phases. Without detailing the analogous arguments, we state the similar result: for $0<\Delta_e\ll |h_y|$ and $L_x$ even, the ideal fixed point Hamiltonian can also be recast as an ordinary TC.

\section{Ground State degeneracy for gapped phases of Eq.~(\ref{Eqn:Decoupled})}\label{Appendix:GSDSpinLattice}

In this Section we provide arguments for the non-trivial size dependence of the GSD for AI$_{i}$ and KW phases, and derive their GSD using direct counting arguments in the original spin representation of the models and without using the fermion mapping. For this purpose we shall set $\delta=0$ in the underlying spin lattice. As we shall see, solutions in the underlying spin lattice confirms the conclusions in Section~\ref{Sec:TopologicalSuperconductor}.~\cite{Wen2009}

First, we consider the trivial phase (AI$_0$), whose ideal fixed point Hamiltonian corresponds to choosing $h_x=h_y=0$ and $h_z>0$ in Eq.~(\ref{Eqn:Decoupled}). The commuting operators are $\Gamma^\varepsilon$ and $\Gamma^e$ which satisfy the constraint in Eq.~(\ref{Eqn:Constraints2}) and the system is equivalent to the standard Toric Code. The GSD is $4$ labeled by $T_{x,y}$ given by Eq.~(\ref{Eqn:t'Hooft1}). They are raised by the Wilson operators $W_{y,x}$ which are products of $Z$ lines across the Torus along two directions as in the Toric Code. For $h_z<0$ (phase AI$_1$), the state $\Gamma^\varepsilon_p=-1$ on each plaquette minimizes the energy. However, for $L_x, L_y$ both odd, this state is forbidden by the total parity constraint in Eq.~(\ref{Eqn:Constraints2}):
\be
\prod_p \Gamma^\varepsilon_p =(-1)^{L_xL_y}=-1.
\ee
Thus the lowest energy state has $\Gamma^\varepsilon=1$ on one plaquette and has a very large degeneracy (these states are not global ground states).

We now turn to the case of $h_z=h_y=0$ and $h_x>0$ in Eq.~(\ref{Eqn:Decoupled}), which corresponds to the ideal fixed point Hamiltonian [see Eq.~(\ref{Eqn:IdealHamiltonian}) and Table~\ref{Table:IdealOperators}] for the phase KW$_{x,0}$ describing a stack of Kitaev wires along the $x$-direction. To determine ground state topological degeneracies, we find all constraints relating operators that enter in Eq.~(\ref{Eqn:IdealHamiltonian}), which in this case are $\Gamma^e_v$ and $U_{x,p}$ in Table~\ref{Table:IdealOperators}. Interestingly, we find that these operators satisfy different global constraints depending on whether $L_y$ is even or odd. For $L_y$ even, their constraint is:
\be\label{Eqn:ConstraintsEven}
\bigg(\prod_{\text{odd rows}} \Gamma^e_v \bigg)\bigg(\prod_{\text{lattice}} U_{x,p} \bigg) =1,
\ee
where the first product is taken over the lower vertices of squares of odd rows. Eqs.~(\ref{Eqn:Constraints2}) and (\ref{Eqn:ConstraintsEven}) give two constraints [note if one sets $U_x=\pm1$ in Eq.~(\ref{Eqn:ConstraintsEven}) and substitutes Eq.~(\ref{Eqn:eConstraint}), we obtain Eq.~(\ref{Eqn:EffectiveTCConstraints}) as it should]. This means that specifying the eigenvalues of all $\Gamma^e_v$ and $U_{x,p}$ (there are $2^{2L_xL_y-2}$ of them in total) still leaves $4$ degrees of freedom in the total Hilbert space. Thus, the ground state is $4$-fold degenerate labeled by topological operators $T_x$ and $T_y$ in Eq.~(\ref{Eqn:t'Hooft1}), which together with $\Gamma^e_v$ and $U_{x,p}$ spans the entire Hilbert space.

For $L_y$ odd we find:
\bea\label{Eqn:ConstraintsOdd}
\bigg(\prod_{\text{odd rows}}^{\text{row} L_y-2} \Gamma^e_v \bigg)\bigg(\prod_{\text{lattice}} U_{x,p} \bigg) =- T_x.
\eea
In contrast to the $L_y$ even case, the constraint relates $T_x$ to $\Gamma^e_v$ and $U_{x,p}$ operators in the Hamiltonian. Therefore, $T_x$ eigenvalue cannot be assigned arbitrarily, and the ground state degeneracy is two labeled by $T_y$ only raised by $W_x$. For example, for $h_x>0$ (phase KW$_{x,0}$) $U_{x,p}=\Gamma^e_v=1$ in the ground state and Eq.~(\ref{Eqn:ConstraintsOdd}) gives $T_x=-1$: periodic boundary condition along the $x$-direction is forbidden. For $h_x<0$ (phase KW$_{x,1}$), $U_{x,p}=-1, \Gamma^e_v=1$ in the ground state, and Eq.~(\ref{Eqn:ConstraintsOdd}) leads to:
\be
T_x = -(-1)^{L_xL_y}.
\ee 
Thus, for $L_y$ odd, the $x$-direction boundary condition is anti-periodic for $L_x$ even and periodic for $L_x$ odd. As is shown in Section~\ref{Sec:PhaseDiagram}, there is a critical point $h_x=h_z$ separating the two phases studied above. Note that Eq.~(\ref{Eqn:ConstraintsOdd}) is not invariant under translation along the $y$-direction by unity, which is a manifestation of `weak symmetry breaking' discussed in the main text.

The degrees of freedom counting for KW$_x$ phases can be summarized as:

\begin{center}
	\begin{tabular}{|c|c|}
		\hline
		operators & degrees of freedom \\ \hline
		$\Gamma^e_v$& $L_xL_y-1$ \\\hline
		$U_x$~&$L_xL_y -z$ \\ \hline
		$T_x$&$z$ \\ \hline
		$T_y$&$1$\\ \hline
	\end{tabular}
	~~$z = 
	\begin{cases} 1,~L_y ~\text{even}\\
	0,~L_y ~\text{odd}\\
	\end{cases}$
	\smallskip
\end{center} 

The ground state degeneracy is then $2^{1+z}$, where $1+z$ is the number of independent $T_{x,y}$ operators.

\section{Chern Numbers for Model Eq.~(\ref{Eqn:Decoupled1})}\label{Appendix:ChernNumber}

The BdG Hamiltonian Eq.~(\ref{Eqn:Decoupled1}) is diagonalized in momentum space and has the form in Eq.~(\ref{Eqn:BCSHamiltonian}), which can be written as:
\be
H = \bm{\sigma}.\bm{c}(\bm{k}),
\ee
where $\bm{c}(\bm{k})= (\Re \Delta(\bm{k}),-\Im \Delta(\bm{k}),\varepsilon(\bm{k}))$. This defines a unit-vector in $\bm{k}$-space $\bm{n}(\bm{k})=\bm{c}/|\bm{c}|$ and the Chern number is:
\be\label{Eqn:ChernNumber1}
C = \frac{1}{4\pi}\int \bigg(\frac{\p \bm{n}}{\p k_x} \times \frac{\p \bm{n}}{\p k_y}\bigg).\bm{n}~d^2k.
\ee
Evaluating Eq.~(\ref{Eqn:ChernNumber1}) near each gap closing point and adding them gives the result in Eq.~(\ref{Eqn:ChernNumber}). For this purpose, we pick a specific point in the parameter space for each phase. For example, for gapless phases we choose $|h_x|=|h_y|=|h_z|$.

%

\end{document}